\pdfoutput=1
\documentclass[10pt]{article}
\usepackage{amsmath,amsthm,amssymb} 
\usepackage{flexisym} 
\usepackage{breqn}
\usepackage{amsfonts}
\usepackage{lmodern}
\usepackage{avant}
\usepackage{courier}

\usepackage[T1]{fontenc}
\usepackage[a4paper]{geometry}
\usepackage{mathtools}   
\usepackage{float}
\usepackage{appendix}
\usepackage{babel}
\usepackage{stmaryrd}
\usepackage{setspace}
\usepackage[numbers]{natbib}
\PassOptionsToPackage{normalem}{ulem}
\usepackage{ulem}
\usepackage[unicode=true,pdfusetitle,
 bookmarks=true,bookmarksnumbered=false,bookmarksopen=false,
 breaklinks=false,pdfborder={0 0 0},pdfborderstyle={},backref=false,colorlinks=true]
 {hyperref}
\hypersetup{
 pdfborderstyle={},pdfborderstyle={},linkcolor=blue,urlcolor=blue,citecolor=blue}
\usepackage{units}
\usepackage{booktabs}
\usepackage{authblk} 
\usepackage{subcaption}
\allowdisplaybreaks

\renewcommand{\toprule}{\specialrule{0.08em}{0pt}{0pt}}
\renewcommand{\midrule}{\specialrule{0.05em}{0pt}{0pt}}
\renewcommand{\bottomrule}{\specialrule{0.08em}{0pt}{0pt}}

\newcommand{\doubleRule}{
    \specialrule{0.05em}{0pt}{0pt} 
    \specialrule{0.05em}{0pt}{0pt} 
}

\title{Human- vs. AI-generated tests: dimensionality and information accuracy in latent trait evaluation}

\author[1]{Mario Angelelli} 
\author[1]{Morena Oliva} 
\author[1]{Serena Arima}
\author[1]{Enrico Ciavolino} 

\affil[1]{Department of Human and Social Sciences, University of Salento, Lecce, Italy}

\date{}

\begin{document}
\maketitle 

\begin{abstract}  
Artificial Intelligence (AI) and large language models (LLMs) are increasingly used in social and psychological research. Among potential applications, LLMs can be used to generate, customise, or adapt measurement instruments. This study presents a preliminary investigation of AI-generated questionnaires by comparing two ChatGPT-based adaptations of the Body Awareness Questionnaire (BAQ) with the validated human-developed version. The AI instruments were designed with different levels of explicitness in content and instructions on construct facets, and their psychometric properties were assessed using a Bayesian Graded Response Model. Results show that although surface wording between AI and original items was similar, differences emerged in dimensionality and in the distribution of item and test information across latent traits. These findings illustrate the importance of applying statistical measures of accuracy to ensure the validity and interpretability of AI-driven tools. 

\texttt{Keywords}: 
Latent trait models; Bayesian IRT; Generative AI; ChatGPT; Test information.  
\end{abstract} 

\section{Introduction} 
\label{sec: introduction} 

The use of artificial intelligence (AI) in data science for social research is gaining increasing attention \cite{buskirk2020machines}. Different AI tools support a variety of functionalities, including data augmentation of evidence from surveys \cite{buskirk2020machines}, data curation, categorisation, and annotation \cite{gilardi2023chatgpt} for thematic assignment \cite{gardos2024identification}, as well as pattern recognition, clustering, network analysis, and text analysis. Focusing on the latter, textual data are essential in social research for deeper investigations of attitudes and perceptions, and appropriate analysis methods are needed when qualitative and quantitative structures should be extracted from (unstructured) textual data. A substantial stimulus to extracting information and meaningful indicators from textual data is represented by generative AI, especially large language models (LLMs). These types of artificial neural networks include attention mechanisms and appropriate encoding of textual information---i.e., embedding. Chatbots based on LLMs, such as ChatGPT, are widely used in a multi-task setting, which includes text translation and multilingual studies \cite{rathje2024gpt}, summarisation, extraction of analytics and quantitative text analysis (including sentiment analysis), and conceptual analysis. An example where LLMs may be valuable support is also represented by cross-cultural research, where the measurement of social constructs can be affected by linguistic factors. 

While the capabilities of generative AI are well established for the processing of collected data, the production of new measurement tools, such as questionnaires, tests, and surveys, is an emerging opportunity that stresses the role of interactions between human experts and LLMs. Thanks to an appropriate prompt design, LLMs can produce prototypes of measurement instruments by training the algorithms with a large documental corpus, and they are able to contextualise the extracted information, aligning it with the scope of the research expressed in the prompt. 

However, the outputs of LLMs should be carefully analysed since they are not the outcome of a consistent modelling of a conceptually grounded theory. In particular, the well-known limitations of current generative AI, which include the lack of explainability, transparency, and the need for unbiased and fair data, become critical when social constructs are investigated. For example, the lack of explainability may limit the replicability of the results, their generalisability, and, hence, the extraction of usable knowledge to support social intervention based on causal effects. 
Linguistic biases in training data could be reflected in the generated text \cite{rozado2020wide,mehrabi2021survey,rozado2023political}, with the potential effect of reinforcing stereotypical expressions or concepts \cite{Abid_2021} even during the collection of new data. The occurrence of unattainable results \cite{alkaissi2023artificial} entails the need for human supervision to avoid producing measurements based on inaccurate information. Despite the efficiency of LLMs in translation and cultural adaptation of text, the misuse of language-specific wording may affect the LLM performance \cite{babaei2025language} and, in the context of test or scale development, generate misalignment between the original and the translated versions, requiring careful revision by experts. Bias in training data may also be enhanced when new synthetic data are produced and, in turn, used for subsequent training \cite{wyllie2024fairness}. Furthermore, research linked to sensitive topics and psychological aspects calls for compliance with methodological and ethical standards and appropriate use of data to preserve their confidentiality and integrity. These issues raise concerns regarding the validity of AI-generated questionnaires, which could compromise the quality of social indicators. 

This work aims to conduct a preliminary investigation of accuracy criteria and methods for comparing validated instruments for psychological measurement with large language model (LLM)-generated alternatives. In particular, it examines how different information criteria can distinguish human- from AI-generated questionnaires. Such methods can enhance accuracy assessment and guide the adaptation or tailoring of questionnaires by evaluating how well test items capture information about respondents across different scores or latent traits in AI-generated solutions. 

First, data are analysed in terms of reliability; following an exploratory factor analysis, we focus on the dimensionality of the questionnaires in terms of latent traits needed to explain the response pattern to items, with special attention to essential unidimensionality \cite{strout1990new}. Furthermore, we explore specific qualitative and quantitative properties of the tests, namely, the ranking of items based on their characteristics (difficulty, discrimination power) and the information they convey across multiple values of the latent traits. These properties and their mutual relation uncover differences between human- and AI-generated measurements, despite their facet similarity. Results pave the way for methodological advances in the exploration of ambiguity in AI-generated measurements, namely, unknown or hidden sources of epistemic uncertainty that may arise when moving from knowledge-based and theoretically grounded measurement design to black box approaches \cite{angelelli2024representations}. 

The rest of the work is organised as follows: in Section \ref{sec: state of the art}, we provide an overview of the state-of-the-art of the use of generative AI, especially LLMs, in social and psychological research. Item response theory and its generalisations, in particular graded response models, are presented in Section \ref{sec: methods}. The data collection procedure, including the prompt design, is reported in detail in Section \ref{sec: data collection}. Section \ref{sec: results} presents the results of the analyses linked to reliability and exploratory factor analysis, while Section \ref{sec: GRM Information functions and dimensionality} concentrates on the dimensionality and information properties of the three questionnaires. Such results are discussed and interpreted in Section \ref{sec: discussion}. Finally, we draw conclusions and, starting from the limitations of this work, identify future research directions in Section \ref{sec: conclusion}.

\section{State of the art} 
\label{sec: state of the art}  

Given the novelty of LLM-based tools, the research on their implications in social and psychological research is still in its early stages. To better contextualise the motivations that stimulated this work, we provide an overview of AI applications in the measurement of sociopsychological constructs. 

\subsection{Generative AI in social research} 
\label{subsec: generative AI in social research}

While automated reporting represents one of the most common applications of AI-powered chatbots, recent studies show that LLMs are perceived as useful even in the conceptual development phases, as they can support researchers in structuring complex models \cite{rask2024beyond}. Evidence of the high performance of LLMs is being collected for a wide range of methodologies, e.g., qualitative analysis of textual data \cite{xiao2023supporting}. Remarkably, LLMs can provide substantial help to domain experts in producing software through assisted coding to support the ad hoc analysis of specific empirical studies. Given the main objective of generative AI of generating new data, opportunities have been identified for data augmentation of survey data through the integration of heterogeneous data types \cite{buskirk2020machines}. LLMs can be used for data augmentation also in social simulations; in particular, plausible responses may be obtained from the simulation of participants with given personality traits, profiles, and backgrounds \cite{salah2024good}. 

Generative AI models are also impacting multiple psychological domains, such as predictive diagnosis, cognitive analysis of mental processes, and support for treatment pathways \cite{salah2024good}. These advances have the potential to substantially improve the efficiency of the clinical processes and patients' quality of life; however, it is noted that the effectiveness of chatbots in therapeutic processes, especially compared with traditional psychotherapy, is still being studied. In the field of emotion analysis, AI can help recognise emotions and sentiments with satisfactory performance; furthermore, LLMs' translational capabilities allow for multilingual processing of text for psychological constructs supporting emotion analysis \cite{rathje2024gpt}. In the field of social psychology, the ability of AI models to analyse large data sets enables the understanding of complex social phenomena to a greater extent; see, e.g., \cite{salah2023may} and references therein for an overview of the main benefits and limitations of generative AI in this field.  

In line with the scope of this work, we note that psychometric applications of intelligent tools have been recently used for the refinement of items during the questionnaire pretesting phase \cite{olivos2024chatgptest}; while valuable insights have been collected from the chatbot, the role of human interpretation and supervision has been stressed once more. Current generative AI approaches allow multimodal data processing and integration, which means that text can be combined with graphical data (e.g., images) to provide a more complete overview of the construct under examination. For example, Zou et al. (2024) \cite{zou2024pilot} explored the perception of a ChatGPT-generated questionnaire compared to its human-generated and hybrid (human-AI) versions; specifically, the authors used ChatGPT to rephrase a questionnaire and explore gender differences in emotional response. The new questionnaire was compared with the original one, also considering facial expressions' recognition to detect emotions, showing a slight preference toward the artificial version. 

On the other hand, several limitations prevent a larger adoption of LLMs in research and intervention. Indeed, beyond the benefits that generative AI can give to data science in psychological and social sectors, its social impact should not be ignored. The appropriate use of LLM chatbots is widely recognised as a need to avoid the consolidation of existing bias, social inequalities, and discrimination, as well as the spread of disinformation; unfairness, low-quality answers, and privacy and regulatory issues also emerge as critical aspects \cite{baldassarre2023social,wach2023dark} that call for appropriate methods for evaluating the outputs of generative AI. In particular, the quality of AI outcomes can be severely affected by hallucinations \cite{ji2023survey}, namely, generated data that are syntactically correct but express invalid, false, or partial information; without proper supervision and validation, they may seriously affect the quality of outputs when they rely on invalid information that is an artefact of the LLM \cite{alkaissi2023artificial}. 

By contrast, it should be remarked that positive effects of AI at the social level have been identified, e.g., in financial inclusion \cite{subramaniam2024exploring}, in training and education \cite{kim2024adapting} and, more generally, as a means to construct new knowledge based on human-AI collaboration \cite{robertson2024game}. This aspect relies on the appropriate interaction between human and artificial agents; for LLMs, this shifts the focus on the relevance of prompt design.

\subsection{Prompt engineering}
\label{subsec: prompt engineering}

The widespread importance that generative AI is gaining in different societal contexts is fostering the role of \emph{prompt engineering} \cite{knoth2024ai}. The term refers to the analysis of interactions with digital systems, including questions proposed as input to chatbots, which enhance the performance in terms of accuracy, precision, and relevance \cite{korzynski2023artificial}. Prompt engineering allows for better control over outputs, contributes to improving content accuracy, and plays a crucial role in maximising the use of language models, connecting user intent and model comprehension. Since prompt quality directly impacts the quality of generated responses, creating an efficient prompt requires taking into account how the model could interpret the input. In principle, this means considering potential issues in the training data, including biases and language or domain-specific attributes, as well as additional limitations that could influence proper comprehension capabilities. 

The role of prompt design is relevant in this work, as it can affect the perceived difference between expected and actual output from ChatGPT. Indeed, the text generated from a specific prompt can produce results that appear similar, or even equivalent, to the expected output, while the outcome from more generic prompts may be less aligned with the expectation or with a human-generated benchmark. In turn, the principles of prompt design can be linked to cognitive processes underlying the construction of questions, including prompts and items in questionnaires. Indeed, prompt engineering relates to the wider domain of human-AI interactions, which have also been discussed as a form of social constructivism, namely, the creation of knowledge through interactions with the environment. In the present case, the environment includes both individuals and artificial agents. In this sense, LLMs may foster knowledge creation in individuals through a scaffolding process, i.e., acting as a source of knowledge that guides the human learner through its zone of proximal development by means of adequate interactions (including prompts) that adapt over the learning process \cite{vygotsky1978mind}. 



However, when designing prompts for AI, the user's prior knowledge is of substantial importance in contextualising the question posed to the chatbot, as many LLMs are not domain-specific experts. The unknown (black box) process of content generation may incorporate new knowledge that is out of context. Furthermore, attention should be paid to avoid the emergence of additional bias during the evaluation of the LLM's outputs \cite{robertson2024game}. In the creation of measurement instruments, the unintended inclusion of psychological or social constructs may introduce misleading information or affect the underlying structure of a test. 
Such changes are more difficult to identify than manifestly invalid or nonsensical content, such as that resulting from some AI hallucinations, but they can invalidate the quality of the indicators obtained from the analysis of data collected with such automated tools.

\section{Methods} 
\label{sec: methods} 

\subsection{Item-Response Theory (IRT)} 
\label{subsec: IRT} 

Item Response Theory (IRT) provides a valid alternative to score-based methods. The main difference between the two approaches relies on the fact that classical assessments are based on test scores, assuming a correspondence between the observed performance on a specific test and the individual's ability. By contrast,  IRT models estimate the subjects' latent traits, accounting for item characteristics such as difficulty and discriminant power.  
The disentanglement of respondents' and items' characteristics is often referred to as the \emph{invariance property} of IRT, which enhances a proper estimate of respondents' traits. The indicators extracted from IRT estimates can foster comparability with other tests thanks to the separation from the specific test. Moreover, IRT can be highly informative on the items' characteristics, which allows considering the informative contribution of each item as a function of the latent trait, as we will see in the next sections.

IRT and its generalisations represent a fundamental premise to better understand the structure that the test measures and the measurement's validity. However, such advantages are entailed by the invariance property, which could not always be reflected in real data. A proper analysis based on IRT relies on the satisfactory fulfilment of some hypotheses verified from empirical data. The first assumption is \emph{unidimensionality}, i.e., the occurrence of only one latent trait; \emph{local independence} represents a conditional independence condition given the existence of a \emph{unique} latent trait measured by the questionnaire. 

The general approach in IRT is to model the probability distribution of the pattern of responses as a function of the $i$-th respondent's latent trait $\vartheta_{i}$ and a set of item features. The basic Rasch model deals with dichotomous responses, and only the item difficulties $\beta_{j}$ are modelled; for this reason, the Rasch model is referred to as the 1-Parameter Logistic (1PL) model. A two-parameter (2PL) extension also takes into account a parameter to evaluate an item's power in discriminating the respondents. A further extension (3PL) includes an additional parameter that describes the probability of guessing. 

Moving beyond responses with two modalities, polytomous data can be examined by means of extensions of the aforementioned models. Among the main approaches are the Partial Credit Model \cite{masters1982rasch} and the Graded Response Model (GRM; see, e.g., \cite{Samejima_1969}). In this work, we adopt the latter model, as it fits well with ordinal answers with more than two levels.  Starting from the probability of a given pattern of responses, the model estimates individual latent traits, as well as difficulty and discrimination item parameters. 

In more detail, GRM expresses the odds of response probabilities at level \emph{at most} $h$, where $h\in\{1,\dots,H-1\}$ labels the ordinal levels \cite{Samejima_1969}. Let $Y_{ij}$ be the response of the subject $i$ to the $j-$th item of the questionnaire, with $j=1,...,M$ and $i=1,...n$. The response variable is a categorical variable with $H$ possible ordered categories. Letting $\Theta$ denote the latent trait that the test aims at measuring, the probability that the $i$-th individual with latent trait $\theta_{i}$ returns a response not higher than $h$ to the $j$-th item is 
 \begin{equation}
\mathrm{logit}(P^{+}_{h|i,j})=\log \left(\frac{P(Y_{i,j}\leq h|\vartheta)}{P(Y_{ij} > h|\theta)}\right) = -\gamma_{j}\cdot (\vartheta_{i}-\beta_{j,h}), \quad h \in \{1,\dots,H-1\} 
\label{eq: GRM}
\end{equation}
where $\gamma_j$ defines the discrimination power of the item $j$ and $\beta_{j,h}$ is the difficulty of response category $h$ of the item $j$. 
A further reduction in the model complexity comes from a relation that decomposes the overall difficulty into two additive contributions
\begin{equation}
\beta_{j,h}=\beta_{j} + \delta_{h},\quad j\in \{1,\dots,M\}, \, h\in\{1,\dots,H-1\} 
\label{eq: difficulty decomposition}
\end{equation}
where $\beta_j$ denotes the $j$-th item-specific difficulty and $\delta_{h}$ is the $h$-th response level threshold. 
These notions enter the following definition of the GRM model 
\begin{align}
    Y_{ij} & \sim \mathrm{cat}(p_{\cdot|ij}), \nonumber \\ 
    p_{h|ij} & := P^{+}_{h|i,j}-P^{+}_{h-1|i,j}, \nonumber \\ 
    P^{+}_{h|i,j} & := P(Y_{ij}\leq h|\vartheta_{i},\beta_{j,h}) =\frac{1}{1+\exp(\gamma_{j}\cdot(\vartheta_{i}-\beta_{j,h}))} 
    \label{eq: GRM responses}
\end{align} 
where we introduce $P^{+}_{0|i,j}=0$ and $P^{+}_{H|i,j}=1$ by consistency with the definition of $P^{+}_{h|i,j}$ as a cumulative probability. 

\subsection{Information functions} 
\label{subsec: information functions}

The aforementioned advantages provided by IRT models, including GRM, allow deriving \emph{functional} descriptions of the latent traits, including extrapolations for out-of-sample values of $\vartheta$. A fundamental quantity to evaluate information in probabilistic modelling is the Fisher information, namely, the expected value of the observed information $-\sum_{h=1}^{H} p_{h|j}(\vartheta)\cdot\frac{\partial^{2}}{\partial\vartheta^{2}} \ln p_{h|j}(\vartheta)$. The study of \emph{information functions} \cite{Birnbaum1968} has been widely explored in the IRT framework. For the graded response model, the relevant quantity expressing the accuracy of each item in describing and differentiating latent traits is the following \emph{item information function} \cite[p. 15]{nering2011handbook}
\begin{equation}
\mathrm{IIF}_{j}(\vartheta)=\sum_{h=1}^{H} \gamma_{j}^{2}\cdot \frac{\left(P^{+}_{h|j}(\vartheta)\cdot \left(1-P^{+}_{h|j}(\vartheta)\right) - P^{+}_{h-1|j}(\vartheta)\cdot \left( 1-P^{+}_{h-1|j}(\vartheta)\right)\right)^{2}}{p_{h|j}(\vartheta)}. 
    \label{eq: item information function }
\end{equation}
A combination of IIFs provides a global index assessing the overall accuracy measure for information conveyed by the test across the latent trait domain. This leads to the introduction of the \emph{Test Information Function} (TIF; see, e.g., \cite{samajima1994estimation}) 
\begin{equation}
\mathrm{TIF}(\vartheta)=\sum_{j=1}^{M}\mathrm{IIF}_{j}(\vartheta). 
    \label{eq: test information function} 
\end{equation} 
For our purposes, we also examine a normalised version of the IIF 
\begin{equation}
    \mathrm{IIF^{(n)}}_{j}(\vartheta) = C_{j}^{-1}\cdot \mathrm{IIF}_{j}(\vartheta) 
    \label{eq: normalised IIF} 
\end{equation}
with the normalising constant 
\begin{equation}
C_{j}=\int_{\mathcal{D}}\mathrm{IIF}_{j}(\vartheta)d\vartheta  
    \label{eq: normalising constant IIF} 
\end{equation} 
over the domain $\mathcal{D}$ of latent traits. 
It follows that a normalised version of the TIF (\ref{eq: test information function}) is defined as
\begin{equation} 
    \mathrm{TIF^{(n)}} = \frac{1}{M}\cdot \sum_{j=1}^{M} \mathrm{IIF^{(n)}}_{j}(\vartheta). 
    \label{eq: weighted test information function}
\end{equation} 

GRM and IIFs have been proven useful in the analysis of scales investigating quality of life and mental health \cite{uttaro1999graded}. In this work, these notions will be adapted to explore analogies and divergences between human- and AI-generated questionnaires.  

\section{Data collection} 
\label{sec: data collection} 

This work explores different interpretations of the items in the Body Awareness Questionnaire (BAQ; see \cite{shields1989body}). While an initial factor analysis of this questionnaire highlighted four sub-factors, subsequent studies identified this measurement as unidimensional in relation to interoception \cite{ferentzi2021examining}. In the following section, we further discuss the dimensionality of BAQ, highlighting the test coherence in latent trait modelling. 

\subsection{Questionnaire} 
\label{subsec: questionnaire} 

The BAQ version adopted in this study is composed of $M = 18$ items measured on a Likert scale with $H = 7$ points, where 1 = ``\textit{Not at all true for me}'' and 7 = ``\textit{Very true for me.}'' The questionnaire aims at measuring the awareness and attention degrees of bodily reactions that are not primarily due to mental or emotional states; in particular, the questionnaire examines bodily cycles and rhythms, small functionality changes, and anticipation of reactions. 

\subsection{Prompt definition} 
\label{subsec:prompt definition} 

Taking the BAQ as a reference, several prompts were formulated, and among them the one with the highest specification and the one, on the other hand, with almost no specification were chosen. 

With these results in mind, ChatGPT was asked to recreate a test that was similar to the original BAQ. The prompts and chats used varied to test and limit the effects of the influence of previous interactions. The first prompt has a very high degree of specification, in which examples of possible questions are also given based on those found in the literature. The resulting questionnaire, GPT.v1, consists of items that are in one-to-one correspondence with the ones in the original BAQ: GPT.v1 is obtained by rephrasing/translating the original BAQ items.
Indeed, the created items were ordered as the original ones. 

The second questionnaire (GPT.v2) was created by defining a less specific prompt: in particular, ChatCGP was asked to produce a psychological test with 18 items on a 7-point Likert scale aimed at evaluating the attention to bodily reactions not influenced by emotive status. 


\subsection{Data collection}
\label{subsec: target sample and data collection}
The questionnaire explores facets of body awareness related to noticing responses or changes in body processes, predicting body reactions, sleep-wake cycles, and disease onset. 

Following the definition of the three versions of the questionnaire to be administered, three corresponding Google Forms were created and shared online. Participation was voluntary, and informed consent was obtained from all respondents prior to accessing and completing the questionnaire. The informed consent explained the objectives of the research and guaranteed that no sensitive personal data would be collected or disclosed. No ethical concerns arose since the study involved only anonymised questionnaire responses and collected no sensitive or personally identifiable information. In this respect,
$n=178$ respondents were randomly assigned to the three questionnaires. Specifically, $n_{BAQ}=56$ subjects 
answered questions in the original BAQ; $n_{GPT.v1}=57$ subjects 
answered questions generated by ChatGPT with the most specific prompt; $n_{GPT.v2}=65$ subjects 
answered questions generated by ChatGPT with the most generic prompt. 
The respondents are homogeneous in terms of potential confounding factors such as reported age and province of residence. The simple sampling design is adequate in this pilot study context, as the current research is focused on investigating potential differences in item characteristics rather than respondents' latent trait quantification.



The data collection lasted two weeks. After the forms were closed, the responses were analysed by using \texttt{R}.

\section{Results: reliability and EFA}
\label{sec: results}

\begin{figure}[!ht]
    \centering
    \includegraphics[width=\linewidth]{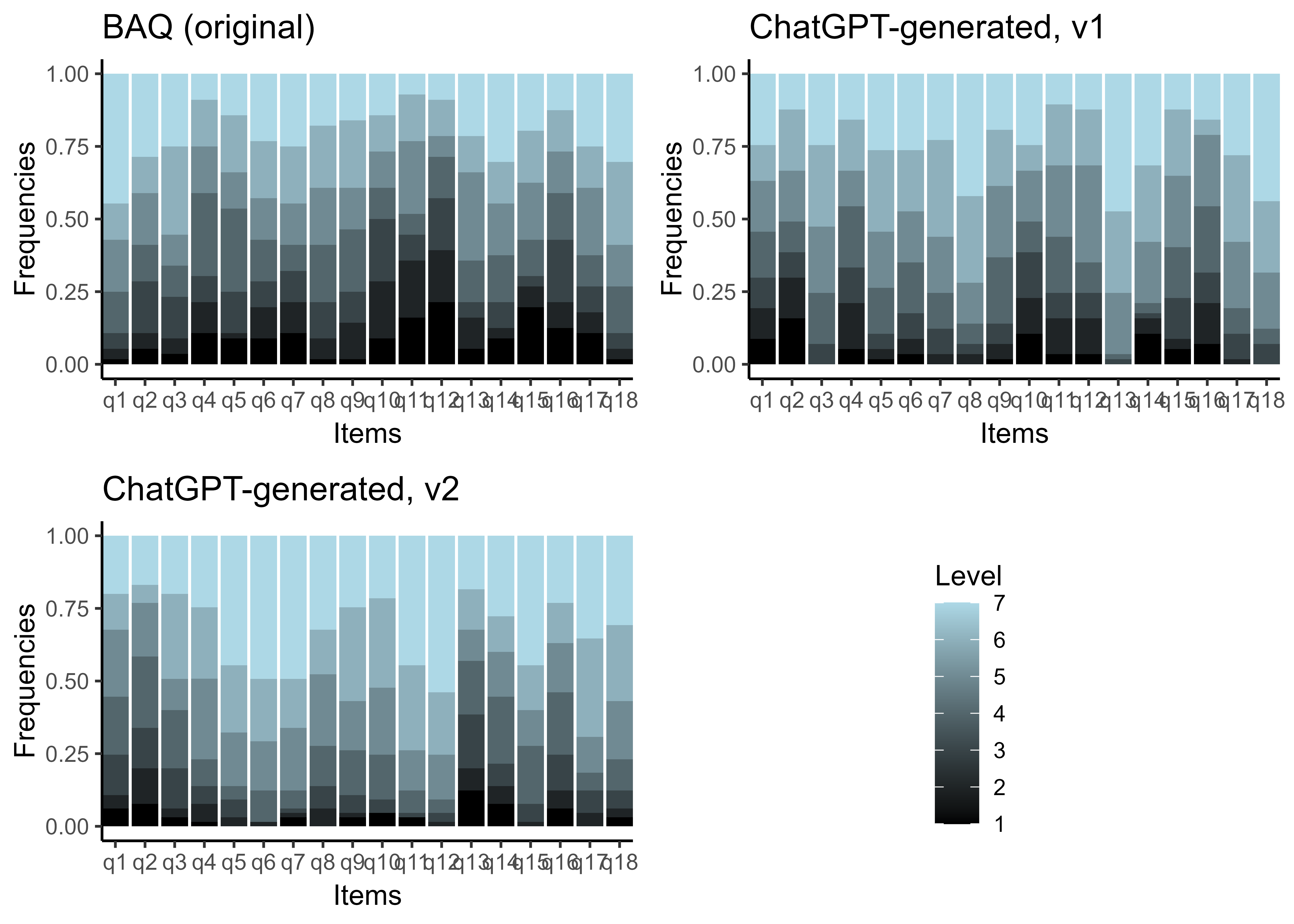}
    \caption{Frequency distribution of response levels for each item in the original BAQ questionnaire (upper-left) and the ChatGPT-generated version v1 (upper-right) and v2 (lower-left).}
    \label{fig: frequency distributions} 
\end{figure}

In Figure~\ref{fig: frequency distributions}, we present a graphical depiction of the frequency distribution of response levels. Then, we move to the examination of whether \emph{reliability} can uncover differences between the three versions of the questionnaire. 

\subsection{Reliability 
} 
\label{subsec: reliability} 

The assessment of the questionnaires' reliability informs us about their internal consistency, but they also provide us with insights about their internal dimensional structure.  

First, focusing on potential differences among the three measurements, we concentrated on Cronbach's $\alpha$ \cite{cronbach1951coefficient} for each test to compare them to assess significant differences. For this purpose, we adopted the two-sided statistical test designed by Feldt (1980) \cite{feldt1980test} and specified by Charter and Feldt (1996) \cite{charter1996testing}. All the pairwise comparisons between Cronbach's $\alpha$ for the different questionnaires were non-significant at significance level $5\%$: specifically, we found p-values equal to $0.212$, $0.616$, and $0.429$ when comparing, respectively, the original BAQ and GPT.v1, the original BAQ and GPT.v2, and GPT.v1 and GPT.v2. 
Therefore, the questionnaires do not differ in terms of reliability measured in terms of Cronbach's $\alpha$. 

Moving to a more detailed analysis of the different aspects of test consistency, we evaluated multiple indices that measure different aspects of the tests' reliability. The classical Cronbach's $\alpha$ is evaluated from numerical scores, which may constitute an approximation of the ordinal data collected from the questionnaires; however, a more appropriate reliability measure that takes into account the qualitative nature of the data is the \emph{ordinal} $\alpha$ \cite{gadermann2012estimating,gadermann2024ordinal}. Along with Cronbach's $\alpha$ and its ordinal counterpart, we also report the $\omega$ coefficient and the composite reliability $\varrho_{C}$. Two types of $\omega$ coefficients are reported, namely, the ``conditional'' $\omega$ and the hierarchical $\omega$ \cite{mcdonald2013test}; both of them take into account the variance explained by a latent factor, but the former relates it to the model-implied covariance, while the latter uses the empirical (observed) covariance matrix. Contrary to Cronbach's $\alpha$, the composite reliability $\varrho_{C}$ takes into account weighted contributions from the different items, which provides a more appropriate reliability estimate even in more complex measurement models, such as Structural Equation Modelling (SEM); see, e.g., \cite{hair2020assessing}, p. 104. 

In Table \ref{tab: reliability indices}, we show the estimates of such reliability coefficients for the three questionnaires under examination. Percentile bootstrap with $R=1{,}000$ replications was conducted to obtain two-sided $95\%$ confidence intervals (CIs) for all these coefficients. 

\begin{table}[!ht]
\centering
\caption{Cronbach's $\alpha$, $\omega$ coefficients, and composite reliability $\varrho_{C}$ for the three questionnaires, along with their $95\%$ bootstrap CIs.}
\label{tab: reliability indices}
\begin{tabular}{clll}
\toprule
 & \multicolumn{1}{c}{\textbf{BAQ (orig.)}} & \multicolumn{1}{c}{\textbf{GPT.v1}} & \multicolumn{1}{c}{\textbf{GPT.v2}} \\
\midrule
$\alpha$ & 0.839 [0.763; 0.885] & 0.775 [0.664; 0.838] & 0.815 [0.685; 0.876] \\
$\alpha_{\mathrm{ord}}$ & 0.848 [0.775; 0.891] & 0.825 [0.743; 0.872] & 0.864 [0.777; 0.908] \\
$\omega$ & 0.848 [0.715; 0.894] & 0.795 [0.691; 0.857] & 0.845 [0.734; 0.900] \\
$\omega_{3}$ & 0.893 [0.636; 0.963] & 0.813 [0.639; 0.911] & 0.882 [0.720; 0.976] \\
$\rho_{C}$ & 0.865 [0.801; 0.908] & 0.843 [0.767; 0.892] & 0.876 [0.798; 0.922] \\
\bottomrule
\end{tabular}
\end{table} 
It is stressed that the second (unconditional) $\omega_{2}$ coefficient is not reported, as it coincides with $\omega$ in the bootstrap samples. These two coefficients compare the explained variance and could differ when controlling for other factors not included in the analysis, which may suggest the occurrence of a more complex test structure. The equality of such coefficients is expected under the specified single-factor model, which is suited to the IRT modelling. 

Despite being related, reliability (or internal consistency) does not identify with the notion of dimensionality \cite{davenport2015reliability}. In the following subsection, we present a more detailed comparison of the three test structures, including their dimensional and informative aspects, in line with the aim of this work. 

\subsection{Exploratory Factor Analysis} 
\label{subsec: exploratory factor analysis}

To explore the dimensionality issue, we performed Exploratory Factor Analysis (EFA) for assessing dimensionality. Specifically, we considered polychoric correlations suited to the ordinal nature of the data; the eigenvalues of the corresponding correlation matrices were used to evaluate an appropriate number of factors based on the empirical Kaiser criterion \cite{braeken2017empirical} implemented in the \texttt{R} package \texttt{EFA.dimensions}. This method generically shows comparable or higher performance compared to parallel analysis. Due to its sample-based approach, this approach also allows assessing the suitability of the sample size to the dimensional analysis \cite[p. 457]{braeken2017empirical}; in our case, all three sample sizes ($n_{BAQ}=56$, $n_{GPT.v1}=57$, and $n_{GPT.v2}=65$) fulfil the criterion to properly carry out the analysis based on the empirical Kaiser criterion for checking unidimensionality. 

In Table \ref{tab: EFA-empKC}, we show the results of the EFA. 
\begin{table}[!ht]
\centering
\caption{Results from the EFA: empirical Kaiser criterion showing the eigenvalue comparison based on the polychoric correlation matrices for the three questionnaires. The number of factors or dimensions to retain is based on eigenvalues higher than the corresponding reference eigenvalues; see \cite{braeken2017empirical}, Eq. (2). The eigenvalues satisfying this criterion are highlighted in bold.}
\label{tab: EFA-empKC}
\begin{tabular}{lllllll} 
\toprule 
\multicolumn{1}{c}{\textbf{dim}} & \multicolumn{2}{c}{\textbf{Original BAQ}} & \multicolumn{2}{c}{\textbf{ChatGPT.v1}} & \multicolumn{2}{c}{\textbf{ChatGPT.v2}} \\ 
\doubleRule \\ 
1                                & \textbf{5.444}           & 2.455          & \textbf{5.245}          & 2.440         & \textbf{5.841}          & 2.329         \\
2  & 2.179 & 2.245 & 1.945 & 2.233 & \textbf{2.198} & 2.147 \\
3  & 1.639 & 2.041 & 1.695 & 2.032 & 1.627          & 1.969 \\
4  & 1.395 & 1.843 & 1.431 & 1.837 & 1.252          & 1.794 \\
5  & 1.134 & 1.651 & 1.134 & 1.648 & 1.125          & 1.624 \\
6  & 1.121 & 1.466 & 1.040 & 1.466 & 0.933          & 1.458 \\
7  & 1.062 & 1.289 & 0.922 & 1.290 & 0.805          & 1.296 \\
8  & 0.725 & 1.118 & 0.841 & 1.121 & 0.723          & 1.140 \\
9  & 0.664 & 1.000 & 0.840 & 1.000 & 0.610          & 1.000 \\
10 & 0.538 & 1.000 & 0.633 & 1.000 & 0.522          & 1.000 \\
11 & 0.505 & 1.000 & 0.512 & 1.000 & 0.489          & 1.000 \\
12 & 0.377 & 1.000 & 0.410 & 1.000 & 0.451          & 1.000 \\
13 & 0.347 & 1.000 & 0.405 & 1.000 & 0.355          & 1.000 \\
14 & 0.278 & 1.000 & 0.341 & 1.000 & 0.293          & 1.000 \\
15 & 0.254 & 1.000 & 0.246 & 1.000 & 0.233          & 1.000 \\
16 & 0.186 & 1.000 & 0.225 & 1.000 & 0.212          & 1.000 \\
17 & 0.101 & 1.000 & 0.105 & 1.000 & 0.183          & 1.000 \\
18 & 0.051 & 1.000 & 0.030 & 1.000 & 0.148          & 1.000 \\ 
\bottomrule 
\end{tabular}%
\end{table}

The empirical Kaiser criterion suggests retaining one factor for both the original BAQ and the more specific version of the ChatGPT-generated questionnaire (GPT.v1); however, two factors are retained for the more generic questionnaire (GPT.v2). 
This aspect shows that potential differences between the human-generated and the AI-generated measurement instruments may emerge as a change in their dimensional structure. With the original BAQ as the benchmark, we proceed in estimating the GRM model, keeping in mind this discrepancy in the more generic GPT-generated test. 

Before proceeding, it is also worth stressing that the change in the dimensional structure is a first hint of a difference between human-generated and AI-generated measurement instruments; furthermore, the divergence of the GPT.v2 test compared with the more specific GPT.v1 test is evidence of the dependence of the produced test on the prompt provided to the chatbot.

\section{Results: dimensionality and informativeness} 
\label{sec: GRM Information functions and dimensionality} 

\subsection{GRM Bayesian estimation}
\label{subsec: GRM Bayesian estimation} 

We specify the GRM presented in Section \ref{subsec: IRT} in a Bayesian setting: starting from (\ref{eq: GRM responses}), the Bayesian hierarchical model is defined by the specification of the latent traits 
\begin{equation}
    \vartheta_{i}|\gamma,\beta \sim \mathcal{N}(0,1) 
    \label{eq: Bayesian latent traits}
\end{equation}
the discrimination parameters  
\begin{equation}
\log(\gamma_{j}) \sim \mathcal{N}(\mu_{\gamma},\tau_{\gamma}^{-1}), \quad \mu_{\gamma}\sim\mathcal{N}(0,\kappa_{\gamma} 
), \quad \tau_{\gamma}\sim\Gamma(5,b_{\gamma}) 
\label{eq: Bayesian discrimination parameters}
\end{equation}
and the difficulty parameters
\begin{equation}
\beta_{j}\sim\mathcal{N}(\mu_{\beta},\tau_{\beta}^{-1}),\quad 
\mu_{\beta}\sim\mathcal{N}(0,\kappa_{\beta} 
), \quad \tau_{\beta}\sim\Gamma(5,b_{\beta}). 
\label{eq: Bayesian difficulty parameters}
\end{equation}
The response levels' thresholds are described as follows: 
\begin{equation}
    \boldsymbol{\delta}=\mathrm{sort}(\boldsymbol{\hat{\delta}}),\quad \hat{\delta}_{k}\sim\mathcal{N}(0,\tau_{\delta}^{-1}) 
    \label{eq: Bayesian delta parameters}
\end{equation} 
where $\kappa_{\beta} = 100$, $\kappa_{\gamma} = 10$ are real parameters, and $b_{\beta}\sim\Gamma(20,2)$, $b_{\gamma}\sim\Gamma(20,2)$, and $\tau_{\delta}\sim \Gamma(1,5)$ 
come from a $\Gamma$ distribution. 

\begin{figure}[!ht]
    \centering
    \includegraphics[width=\linewidth]{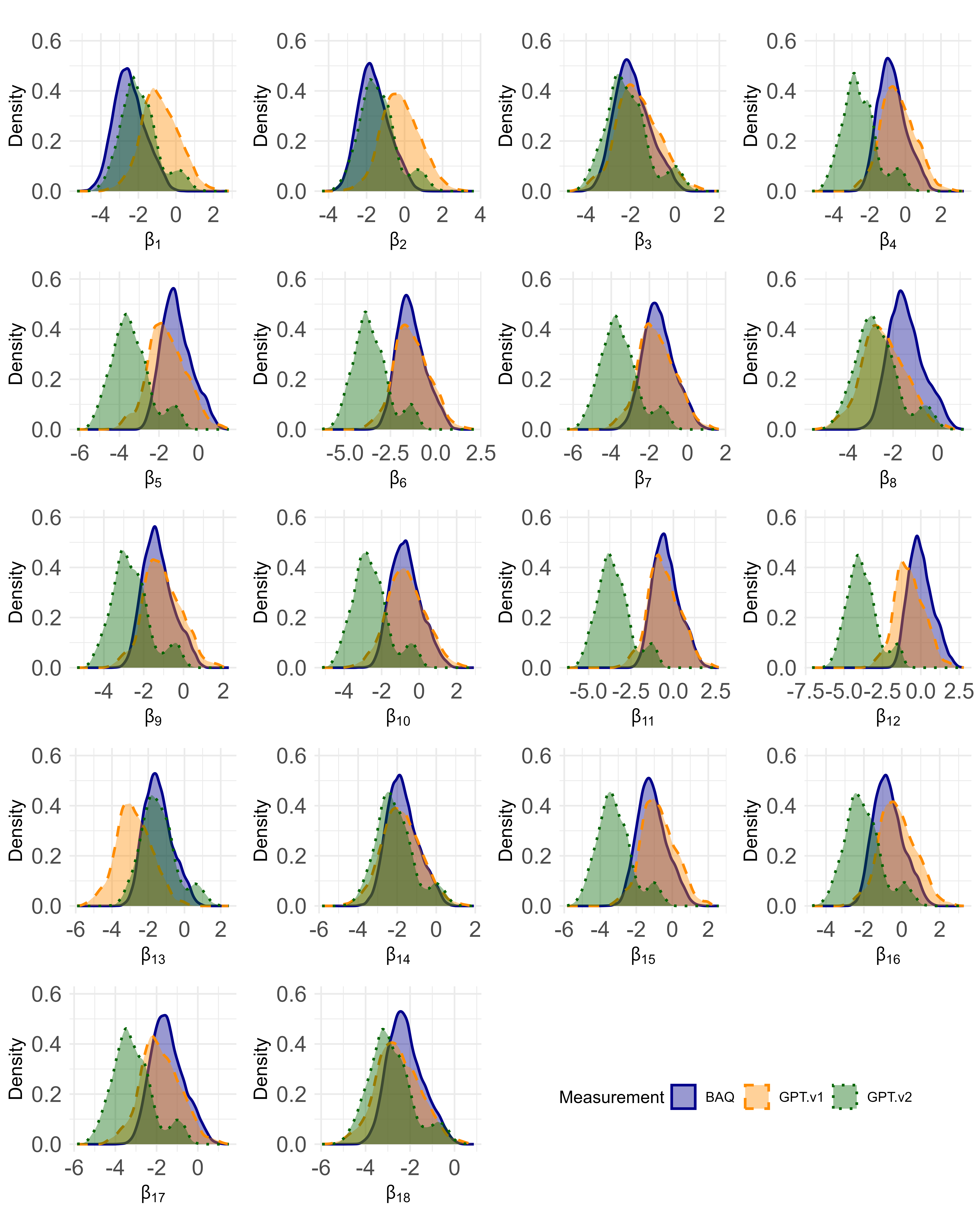}
    \caption{MCMC posterior distribution for $\beta$ difficulty parameters.} 
    \label{fig: beta posterior distributions}
\end{figure}

\begin{figure}[!ht]
    \centering
    \includegraphics[width=\linewidth]{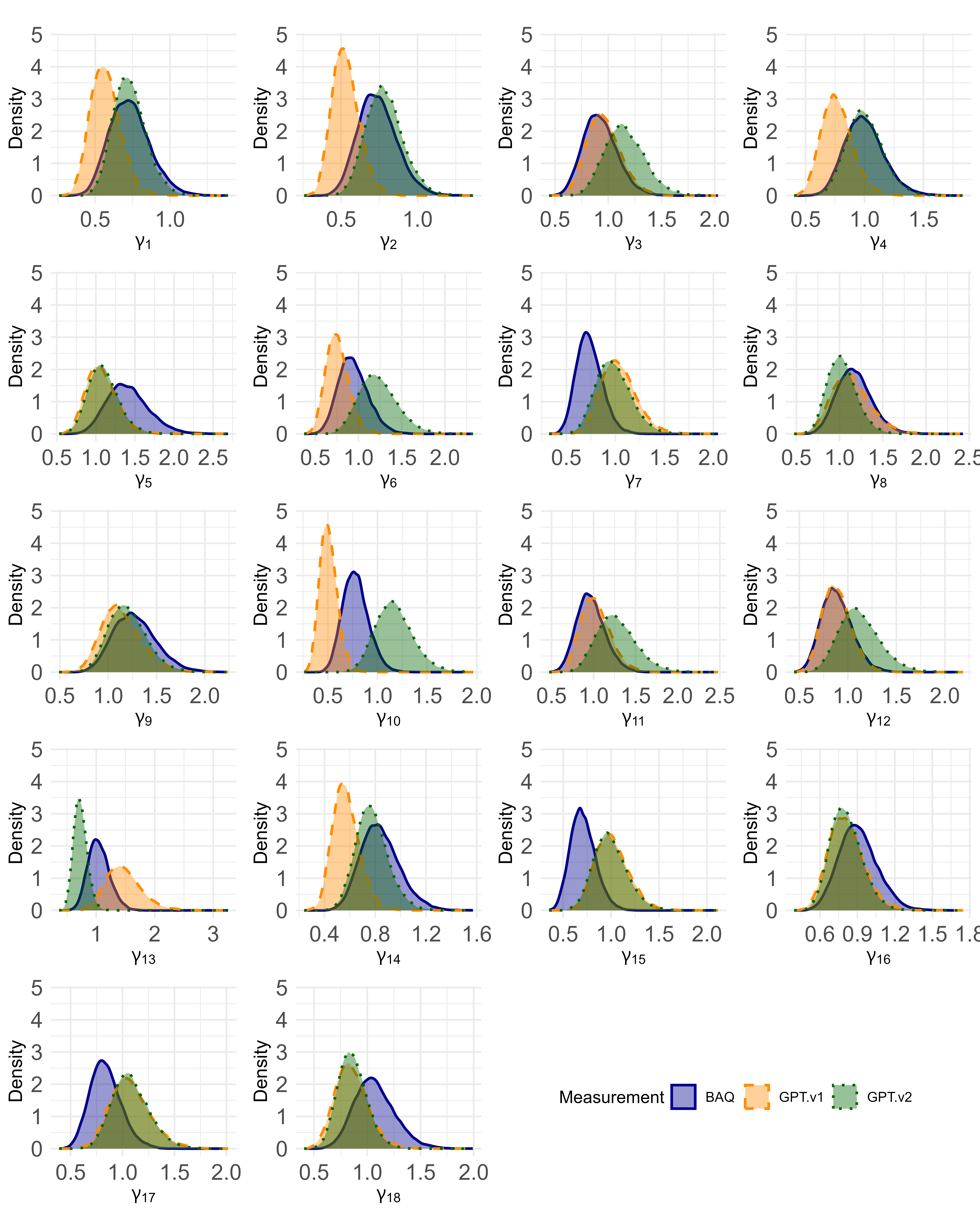}
    \caption{MCMC posterior distribution for $\gamma$ discrimination parameters.} 
    \label{fig: gamma posterior distributions}
\end{figure}

\begin{figure}[!ht]
    \centering
    \includegraphics[width=\linewidth]{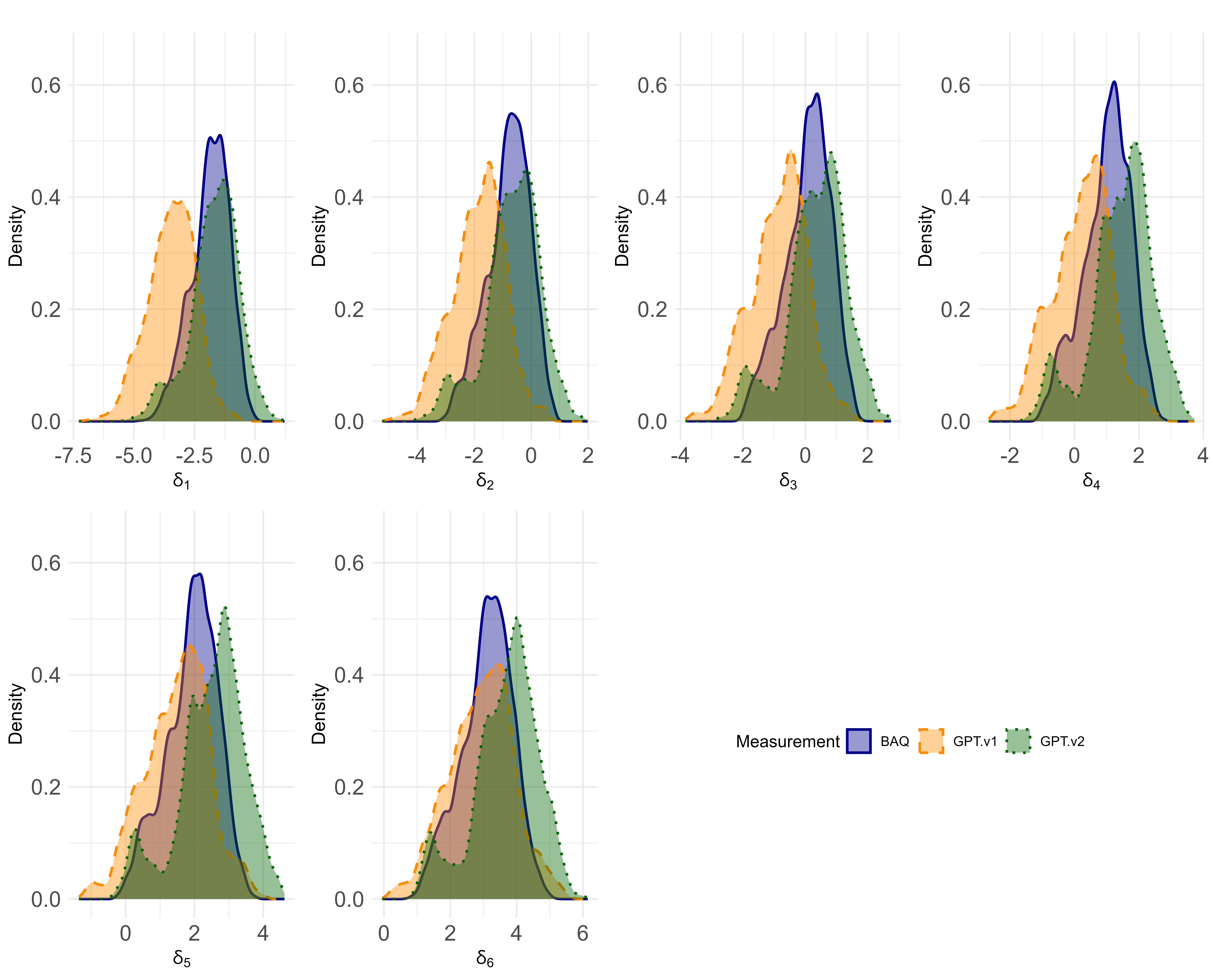}
    \caption{MCMC posterior distribution for $\delta$ level thresholds.} 
    \label{fig: delta posterior distributions}
\end{figure}

Posterior distributions cannot be obtained in closed form. Samples from the posterior distributions are obtained through MCMC machinery: 
we run $n_{\mathrm{MCMC}} = 20{,}000$ samplings with three chains, following an initial burn-in phase composed of $9{,}000$ iterations. 
Focusing on the test characteristics, the summary of the parameters' posterior distributions is presented in the tables 
reported in Appendix \ref{app: summary of GRM estimates from MCMC sampling}. A graphical presentation of the posterior probability distributions is presented in Figures \ref{fig: beta posterior distributions}--\ref{fig: delta posterior distributions}.  


The indices associated with the latent traits can be estimated from the posterior distribution of each $\vartheta$. We choose the median of the posterior distribution $\hat{\vartheta}_{i}$ for the $i$-th respondent to enhance the robustness of these indicators of the latent trait. These estimates are used in the confirmatory analysis for essential unidimensionality discussed in the following. 

\subsection{Essential unidimensionality} 
\label{subsec: essential unidimensionality} 

In line with the scope of this research, the adoption of a graded response model to analyse the collected data requires checking some assumptions underlying the validity of IRT-type modelling. On the one hand, real data hardly fulfil this unidimensionality requirement \cite{zhang2007conditional}; on the other hand, the classical definition of unidimensionality may be a stronger condition than the ones needed for IRT modelling. 

For this reason, a weaker notion of unidimensionality suited to IRT is explored, namely, \emph{essential unidimensionality} \cite{strout1990new}, which formalises the concept of dominant dimensions and essential dimensionality by concentrating on conditional covariance between item pairs given the information on the individuals' latent space. This notion has been exploited by Zhang and Stout (1999) \cite{zhang1999theoretical} to introduce appropriate indices for evaluating essential dimensionality (DETECT), as well as the occurrence of an ``approximate simple structure'' of the test (ASSI and RATIO), which refers to the possibility of partitioning the items consistently with the conditional correlation structure. The construction of these indices relies on the correlation signs as a proxy of membership to the same dimension (or, in the case of multidimensionality, different dimensions) given a \emph{composite}, namely, an individual indicator summarising one or multiple latent traits. The computation of indices from empirical data to evaluate dimensionality and distinguish this information from random noise has been extended to polytomous items through the polyDETECT method \cite{zhang2007conditional}. Therefore, we conduct our confirmatory analysis based on polyDETECT, whose associated computation was carried out by means of the \texttt{sirt} \texttt{R} package. 

In Table \ref{tab: DETECT results}, we report the analyses for two different choices of the latent trait indicator: a naive score based on the median of each respondent's answers and the median $\hat{\vartheta}_{i}$ obtained from the Bayesian GRM estimation. Indeed, the role of conditional correlations in the study of essential unidimensionality gives importance to the definition of a composite and, hence, to the way scoring is attributed to individuals. The criteria to assess essential unidimensionality are DETECT $<0.20$, ASSI $<0.25$, and RATIO $<0.36$ (see, e.g., \cite{jang2007investigation}, p. 7, and \cite{zhang1999theoretical}, Sec. 8). Two different weighting schemes are considered for conditional covariances of item pairs to evaluate such statistics. In both cases, the essential unidimensionality criteria are satisfied when the GRM estimates $\hat{\vartheta}_{i}$ are used. 

\begin{table}[!ht]
\centering
\caption{Results of the essential unidimensionality analysis for both naive estimates of the latent trait score.}
\label{tab: DETECT results}
\begin{tabular}{cllllll} 
\toprule \\ 
 & \multicolumn{2}{c}{\textbf{BAQ}} & \multicolumn{2}{c}{\textbf{GPT.v1}} & \multicolumn{2}{c}{\textbf{GPT.v2}} \\ 
\doubleRule \\ 
 & \multicolumn{6}{c}{\textit{Naive latent trait score (response median values)}}                               \\ 
\doubleRule \\ 
       & Unweighted  & Weighted  & Unweighted  & Weighted  & Unweighted & Weighted \\ 
\midrule \\ 
DETECT & -1.857      & -1.857    & 2.657       & 2.657     & 0.236      & 0.236    \\
ASSI   & -0.033      & -0.033    & 0.085       & 0.085     & -0.163     & -0.163   \\
RATIO  & -0.065      & -0.065    & 0.108       & 0.108     & 0.012      & 0.012    \\ 
\doubleRule \\ 
       & \multicolumn{6}{c}{\textit{GRM latent trait score (median $\theta$ values)}} \\ 
\doubleRule \\ 
       & Unweighted  & Weighted  & Unweighted  & Weighted  & Unweighted & Weighted \\ 
\midrule \\ 
DETECT & -6.077      & -6.077    & -5.487      & -5.487    & -5.679     & -5.679   \\
ASSI   & -0.176      & -0.176    & -0.203      & -0.203    & -0.242     & -0.242   \\
RATIO  & -0.208      & -0.208    & -0.252      & -0.252    & -0.261     & -0.261  \\ 
\bottomrule 
\end{tabular}%
\end{table} 

It is noted that the more specific version, GPT.v1, does not pass the confirmatory test for essential unidimensionality if a \textit{naive} test score is adopted, namely, the median of a respondent's responses. This basic score is suited to the ordinal nature of the collected data, but it does not convey enough information nor distinguish the latent traits' and the items' contributions, as discussed in the previous sections. The composites attributed to latent traits through GRM provide essential information about the respondents' perceptions. 

In the following sections, we move our focus to information contributions about the test structure.

\subsection{Lack of co-monotonicity} 
\label{subsec: lack of co-monotonicity}

The outcome of the GRM estimation can be used to evaluate and rank the items based on different criteria associated with their characteristics. In particular, items can be ordered based on an estimation of their difficulty or discrimination parameters derived from the corresponding posterior distributions. As in the previous section, we use the median values of the $\beta$ and $\gamma$ parameters to enhance the robustness of the analysis. 

The apparent correspondence between the original BAQ and GPT.v1 should be reflected in the analogous ordering of their items based on the magnitude of the associated parameters; in other words, given two indices $j_{1},j_{2}\in\{1,\dots,M\}$, we expect the implication 
\begin{equation*} 
\hat{\beta}_{j_{1}}^{\mathrm{(BAQ)}} < \hat{\beta}_{j_{2}}^{\mathrm{(BAQ)}} \Rightarrow \hat{\beta}_{j_{1}}^{\mathrm{(GPT.v1)}} < \hat{\beta}_{j_{2}}^{\mathrm{(GPT.v1)}}
\end{equation*} 
to hold. We refer to this property as \emph{co-monotonicity}. 

\begin{figure}[!ht]
    \centering
    \includegraphics[width=\linewidth]{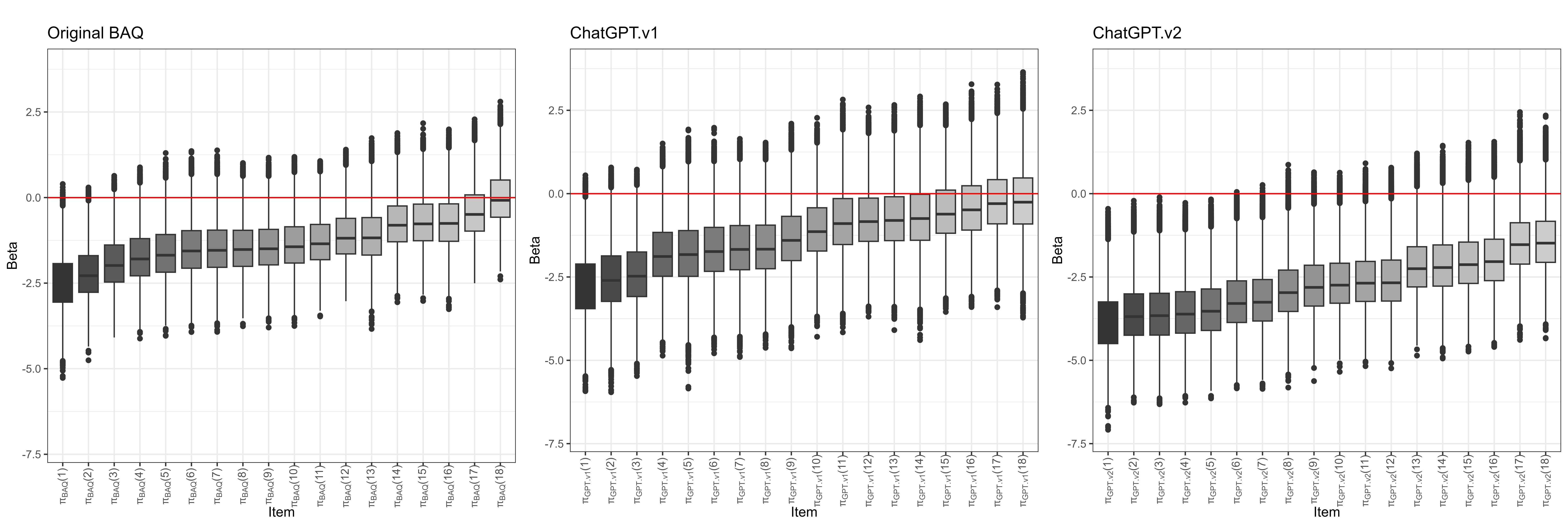}
    \caption{Box plots of MCMC posterior distributions for $\beta$ difficulty parameters. For each test, the parameters have been sorted in increasing order.} 
    \label{fig: sorted boxplot beta posterior distributions}
\end{figure}

\begin{figure}[!ht]
    \centering
    \includegraphics[width=\linewidth]{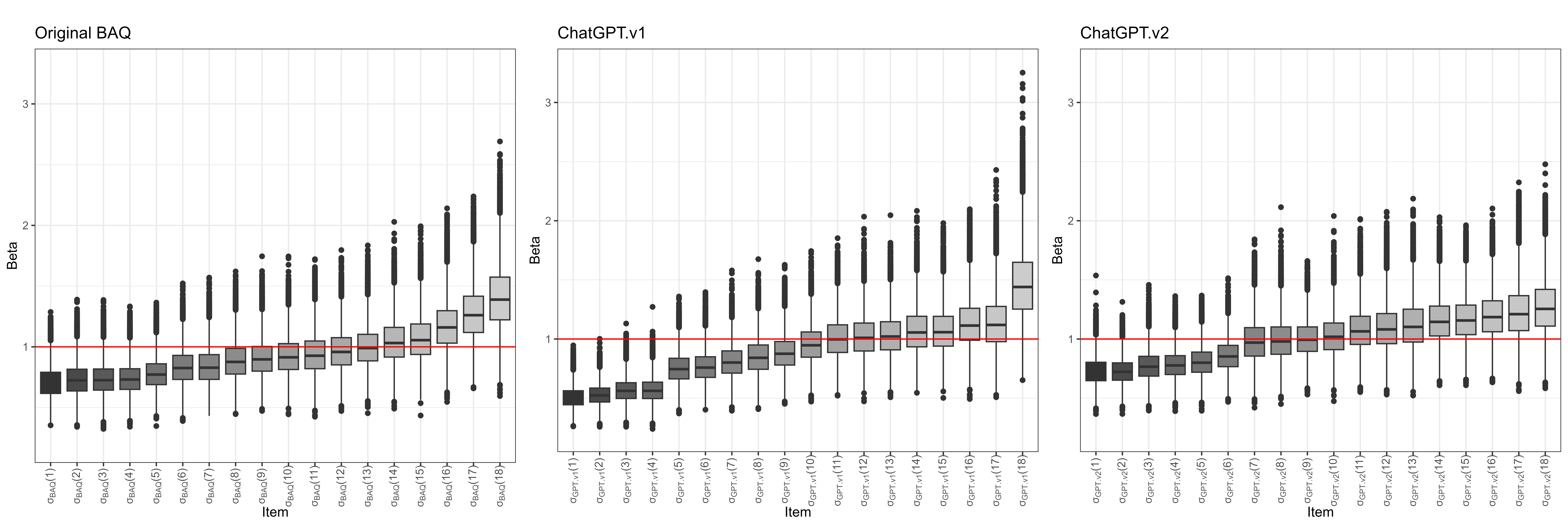}
    \caption{Box plots of MCMC posterior distributions for $\gamma$ discrimination parameters. For each test, the parameters have been sorted in increasing order.}  
    \label{fig: sorted boxplot gamma posterior distributions}
\end{figure}

From the MCMC sampling, we obtain the following orderings expressed by a permutation $\pi$ of the items' label set $\{1,\dots,M\}$ induced by the $\hat{\beta}$ parameters: 
\begin{align} 
\pi_{\mathrm{BAQ}} & = (1, 18, 3, 14, 2, 7, 17, 13, 8, 6, 9, 5, 15, 4, 10, 16, 11, 12), \nonumber \\ 
\pi_{\mathrm{GPT.v1}} & = (13, 18, 8, 17, 14, 3, 7, 5, 6, 9, 1, 15, 12, 10, 11, 4, 16, 2), \nonumber \\ 
\pi_{\mathrm{GPT.v2}} & = (12, 6, 7, 11, 5, 15, 17, 18, 8, 9, 10, 4, 3, 14, 16, 1, 13, 2) 
\label{eq: ordered beta labels by median} 
\end{align} 
Here, $\pi_{\mathrm{BAQ}}$ refers to the permutation obtained from the subgroup $0$ responding to the original BAQ, while $\pi_{\mathrm{GPT.v1}}$ and $\pi_{\mathrm{GPT.v2}}$ express the ordering generated by the subgroups $1$ and $2$ under the administration of the GPT.v1 and GPT.v2, respectively. By the same token, three permutations $\sigma$ of $\{1,\dots,M\}$ are generated by the $\hat{\gamma}$ parameters: 
\begin{align} 
\sigma_{\mathrm{BAQ}} & = (15, 1, 7, 2, 10, 17, 14, 12, 16, 3, 6, 11, 4, 13, 18, 8, 9, 5), \nonumber \\ 
\sigma_{\mathrm{GPT.v1}} & = (10, 2, 1, 14, 6, 4, 16, 18, 12, 3, 15, 11, 7, 5, 17, 9, 8, 13), \nonumber \\ 
\sigma_{\mathrm{GPT.v2}} & = (13, 1, 14, 2, 16, 18, 7, 15, 4, 8, 17, 5, 12, 3, 10, 9, 6, 11). 
\label{eq: ordered gamma labels by median} 
\end{align} 
These orderings are quite stable in terms of parameter estimators; specifically, when we use the mean instead of the median, we only have a slight change in $\pi_{\mathrm{BAQ}}$, i.e., the transposition $(5,15)$. Furthermore, similar results are found for other choices of the constant hyperparameters in the MCMC sampling. In Figures \ref{fig: sorted boxplot beta posterior distributions} and \ref{fig: sorted boxplot gamma posterior distributions} we show the box plots associated with the parameters' posterior distributions sorted by the $\pi$ and $\sigma$ orderings in (\ref{eq: ordered beta labels by median}) and (\ref{eq: ordered gamma labels by median}), respectively. 

\begin{figure}[!ht]
    \centering
    \includegraphics[width=\linewidth]{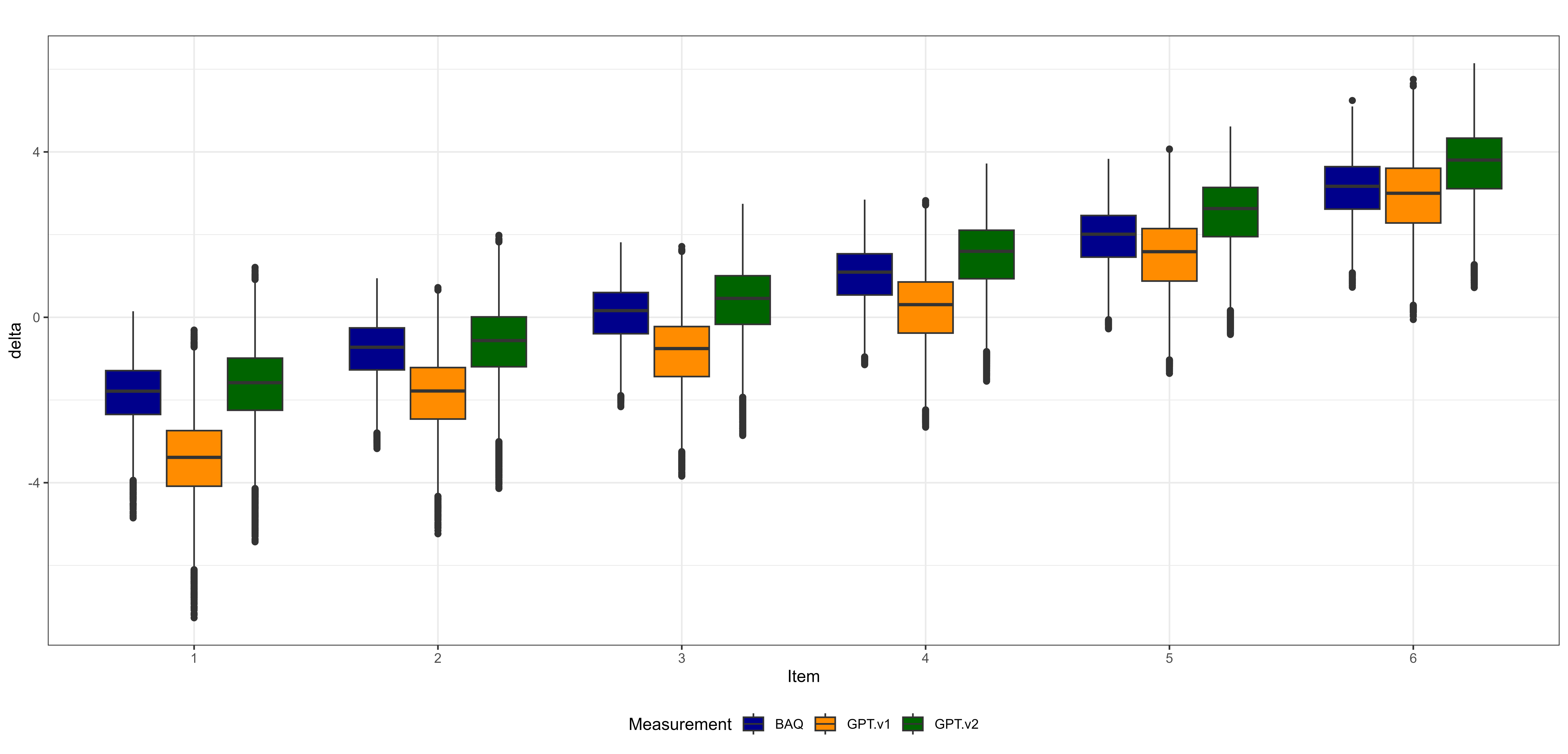}
    \caption{Box plots of MCMC posterior distributions for the $\delta$ response level thresholds.}  
    \label{fig: boxplot delta posterior distributions}
\end{figure}

In addition, Figure \ref{fig: boxplot delta posterior distributions} presents the box plots also for the $\delta$ parameters, which are sorted increasingly by definition to get a classical (i.e., non-negative) distribution starting with the cumulative probabilities $P^{+}_{h|i,j}$ in (\ref{eq: GRM}). 

\begin{figure}[!ht]
    \centering
    \includegraphics[width=\linewidth]{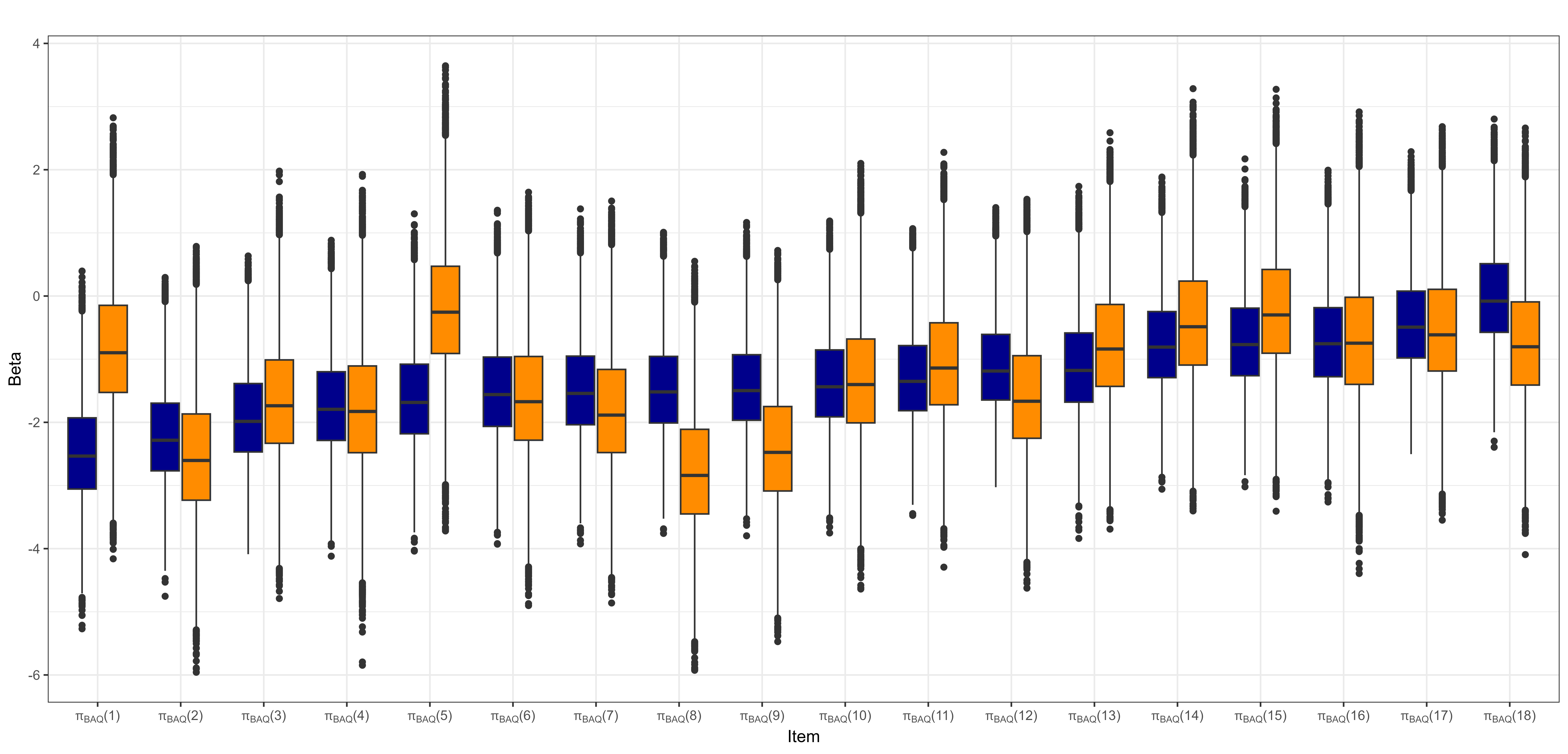}
    \caption{Comparison between difficulty parameters' posterior distributions from the original BAQ and the AI-generated questionnaire GPT.v1. Items are sorted based on the estimate (median) $\hat{\beta}^{\mathrm{(BAQ)}}_{j}$.}  
    \label{fig: BAQ-GPT.v1 boxplot comparison beta}
\end{figure}

\begin{figure}[!ht]
    \centering
    \includegraphics[width=\linewidth]{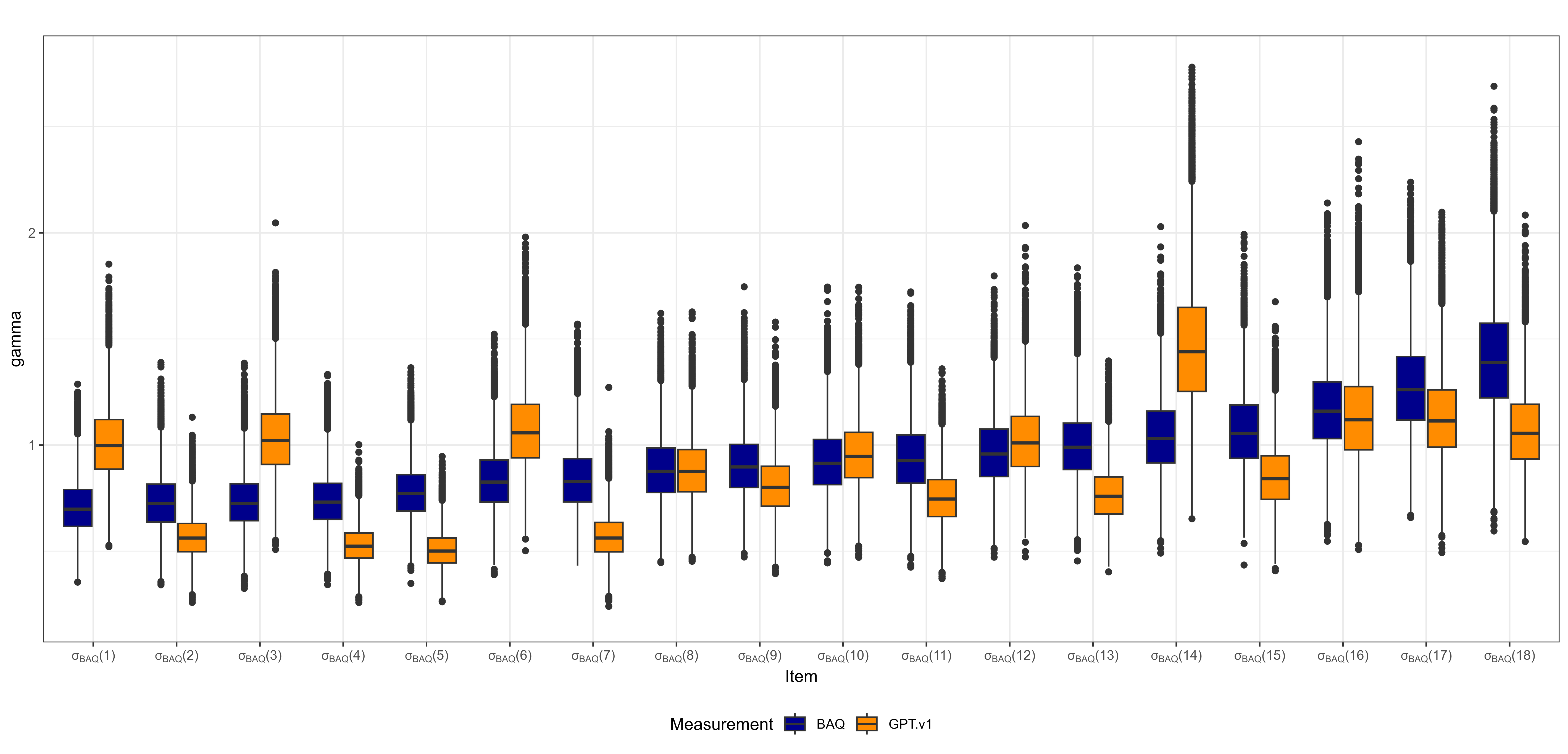}
    \caption{Comparison between discrimination parameters' posterior distributions from the original BAQ and the AI-generated questionnaire GPT.v1. Items are sorted based on the estimate (median) $\hat{\gamma}^{\mathrm{(BAQ)}}_{j}$.}  
    \label{fig: BAQ-GPT.v1 boxplot comparison gamma}
\end{figure}

From (\ref{eq: ordered beta labels by median}) and (\ref{eq: ordered gamma labels by median}), we see that the permutations $\pi_{\mathrm{BAQ}}$ and $\pi_{\mathrm{GPT.v1}}$ differ, as well as $\sigma_{\mathrm{BAQ}}$ and $\sigma_{\mathrm{GPT.v1}}$. This means that co-monotonicity does not hold, which is in contrast with the apparent correspondence between the original BAQ and GPT.v1. A graphical presentation of this aspect is depicted in Figures \ref{fig: BAQ-GPT.v1 boxplot comparison beta} and \ref{fig: BAQ-GPT.v1 boxplot comparison gamma}. 

\subsection{Item and Test Information Function} 
\label{subsec: item and test information functions} 

As the last step of our analysis, we focus on the item and test information functions introduced in Section \ref{sec: methods} since a richer description of the information extracted through the questionnaires across multiple latent traits can support the identification of their mutual similarities and divergences. 

Specifically, a similarity index can be defined in terms of the \emph{overlap} between two functions; dealing with non-negative functions, the overlap is measured as the integral of their minimum 
\begin{equation}
    \cap(I_{1},I_{2}) = \int_{\mathcal{D}} \min\{I_{1}(\vartheta),I_{2}(\vartheta)\} d\vartheta 
    \label{eq: overlap index} 
\end{equation} 
where $\mathcal{D}$ is the common domain of two information functions $I_{1}$ and $I_{2}$. We can specify this similarity index for both $\mathrm{IIF}_{j}$ and $\mathrm{TIF}$ to evaluate an item-wise or test-wise similarity index between the original BAQ and GPT.v1, respectively. To enhance the interpretability of the results, the overlap will be considered in relation to the normalising constants  $C_{1}=\cap(I_{1},I_{1})$ and $C_{2}=\cap(I_{2},I_{2})$ 
\begin{equation}
    \frac{\cap(I_{1},I_{2})}{\sqrt{\cap(I_{1},I_{1})\cdot \cap(I_{2},I_{2})}}. 
    \label{eq: scaled overlap index} 
\end{equation}
We see that $\cap(\mathrm{IIF}_{j},\mathrm{IIF}_{j})$ coincides with the normalising constant $C_{j}$ introduced in (\ref{eq: normalising constant IIF}). A different weighting scheme can be considered by evaluating the overlap operation $\cap$ for the normalised IIFs (\ref{eq: normalised IIF}). 

While overlap assesses the similarity between the two questionnaires, a dominance index can evaluate divergences between them. In analogy with the definition (\ref{eq: overlap index}), we introduce a \emph{dominance} index 
\begin{equation} 
\mathrm{Dm}(I_{1},I_{2})=\int_{\mathcal{D}} I_{1}(\vartheta)\cdot \mathbb{I}(I_{1}(\vartheta)>I_{2}(\vartheta)) d\vartheta
    \label{eq: dominance index} 
\end{equation} 
where $\mathbb{I}$ is the indicator function whose value is $\mathbb{I}(x>y) = 1$ if $x>y$ and $0$ otherwise. In practical terms, $\mathrm{Dm}(I_{1},I_{2})$ is the proportion of the total (normalised) information $I_{1}$ associated with the domain of latent traits described more accurately by $I_{1}$ than by $I_{2}$. To improve the comparison between the original BAQ and GPT.v1, we evaluate the dominance index for normalised IIFs, where $\mathrm{Dm}$ can be interpreted as the proportion of the total information (equal to $1$) in which the $j$-th item of the first questionnaire is more informative on theta than the corresponding item of the second questionnaire. The dominance and overlap indices are not independent, as 
\begin{align} 
    \mathrm{Dm}(I_{1},I_{2})+\mathrm{Dm}(I_{2},I_{1})+\cap(I_{1},I_{2}) & = \int_{\mathcal{D}} \max\{I_{1},I_{2}\} d\vartheta + \int_{\mathcal{D}} \min\{I_{1},I_{2}\} d\vartheta \nonumber \\ 
    & = \int_{\mathcal{D}} I_{1} d\vartheta + \int_{\mathcal{D}} I_{2} d\vartheta 
    \label{eq: relation between overlap and dominance indices } 
\end{align} 
which is equal to $2$ when these information functions refer to normalised IIFs. 

\begin{figure}[!ht]
    \centering
    \includegraphics[width=\linewidth]{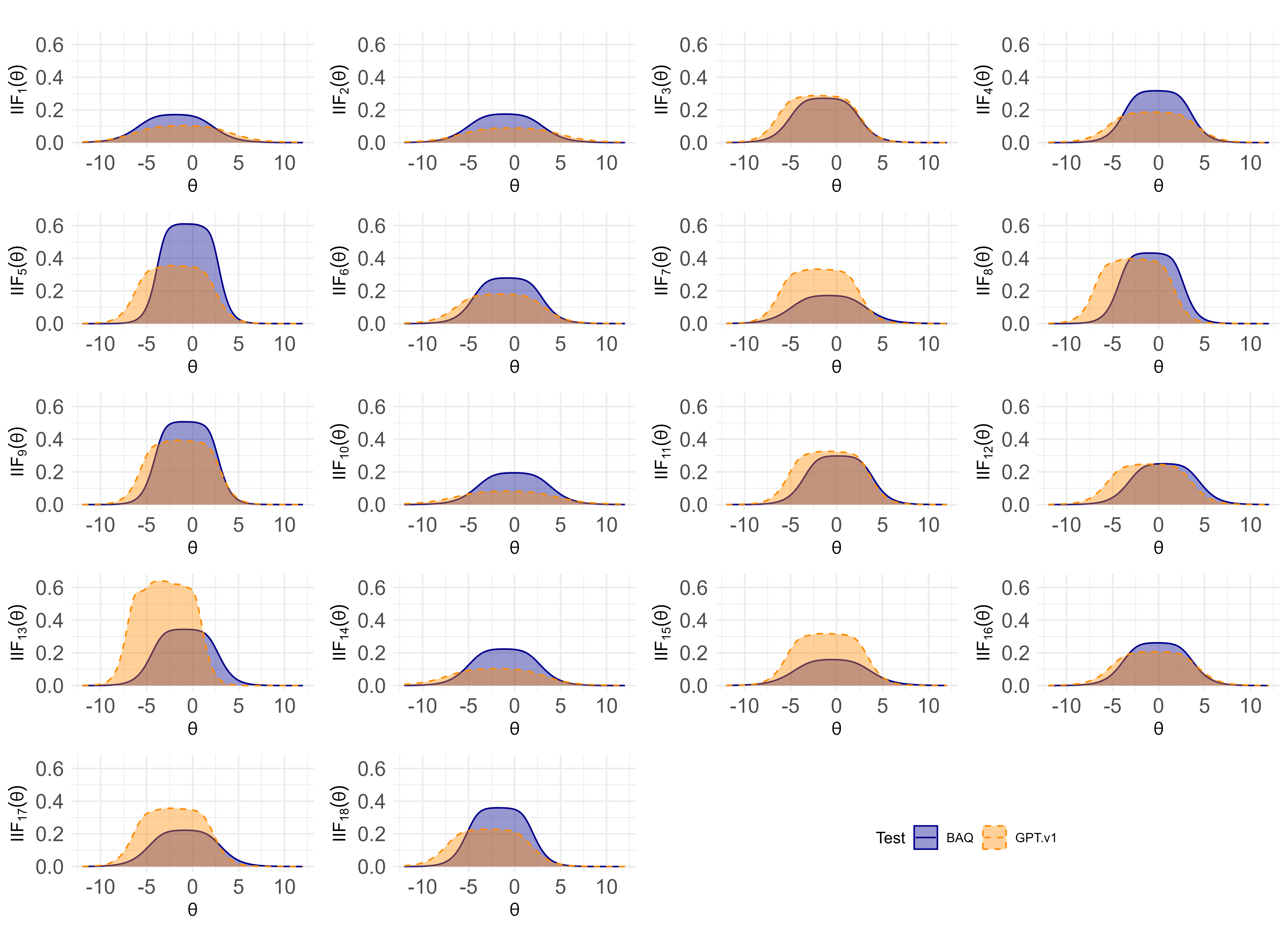}
    \caption{Item Information Functions plots for each item in the original BAQ and the AI-generated GPT.v1 tests.}
    \label{fig: IIF plots}
\end{figure}

\begin{figure}[!ht]
    \centering
    \includegraphics[width=\linewidth]{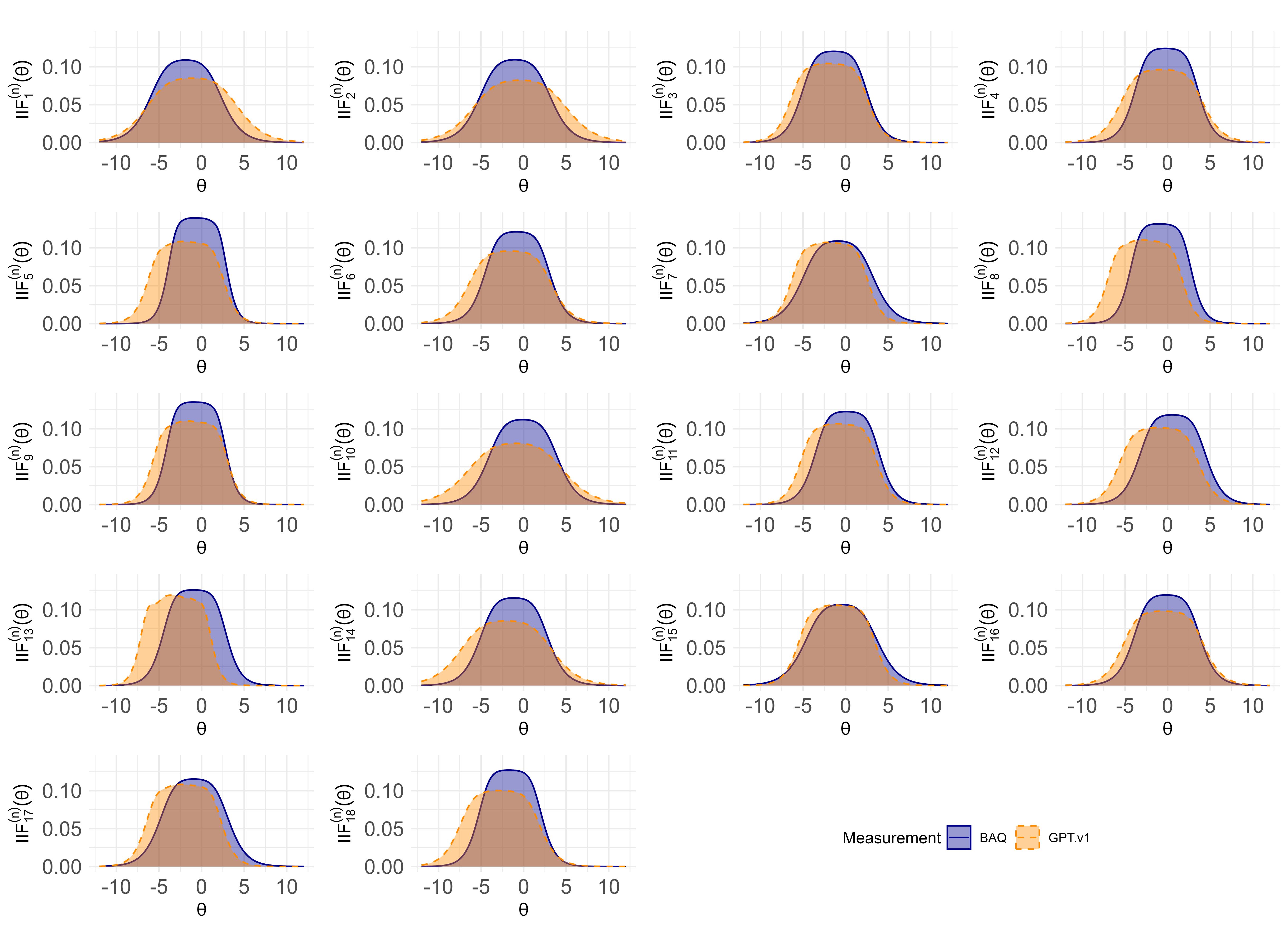}
    \caption{Normalised IIFs plots for each item in the original BAQ and the AI-generated GPT.v1 tests.}
    \label{fig: IIF normalised plots}
\end{figure}

With these premises, the IIFs obtained from the previous procedures are shown in Figure \ref{fig: IIF plots}; even in this case, we use the median of the MCMC posterior distributions as an estimation of the parameters in the information functions.  
The scaled versions of the IIFs according to (\ref{eq: normalised IIF}) are depicted in Figure \ref{fig: IIF normalised plots}. 

\begin{figure}[ht]
   \subfloat[Standard Test Information Function]{\label{fig: TIF plot}
      \includegraphics[width=.45\textwidth]{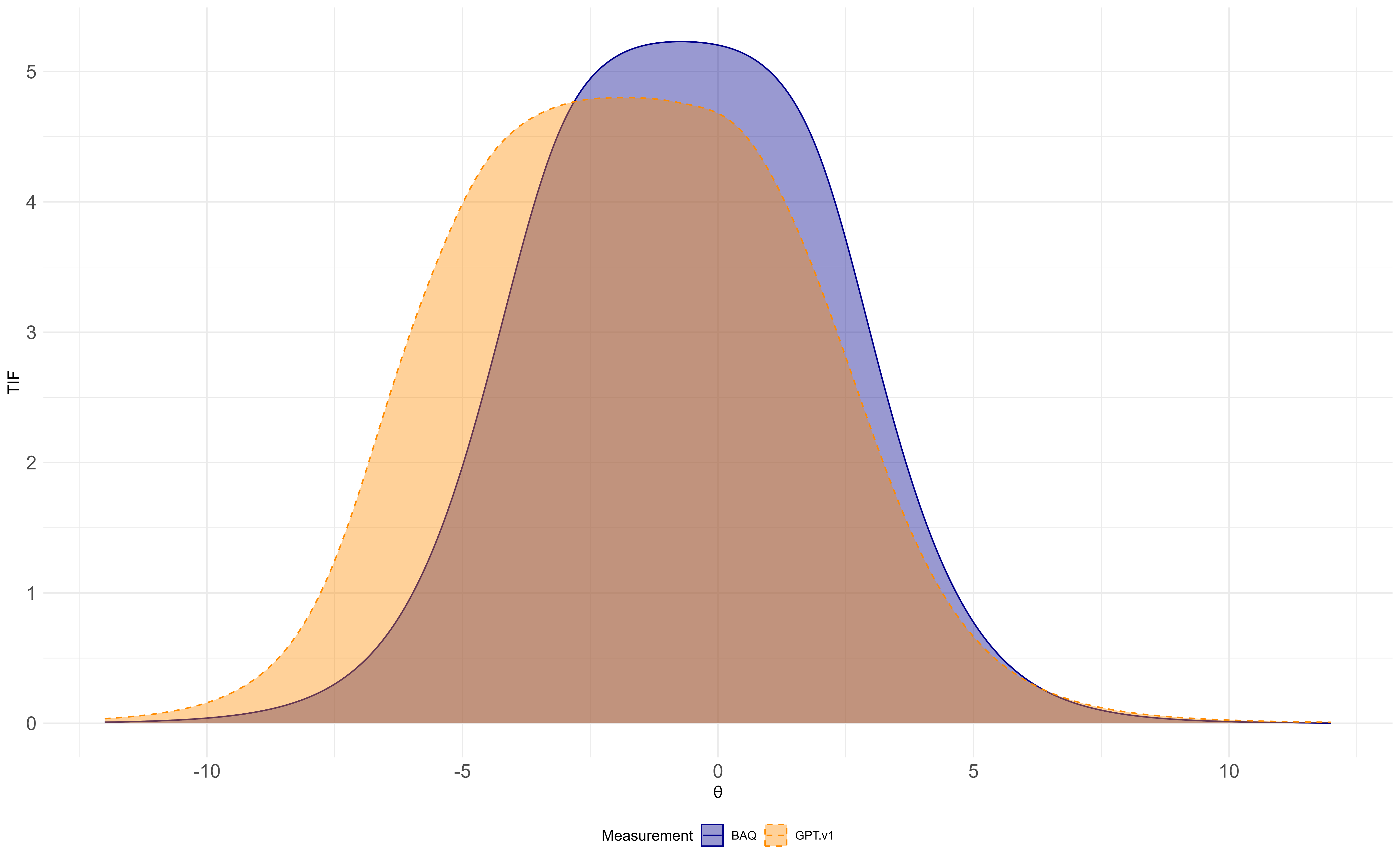}} 
~
   \subfloat[Normalised Test Information Function]{\label{fig: wTIF plot}
      \includegraphics[width=.45\textwidth]{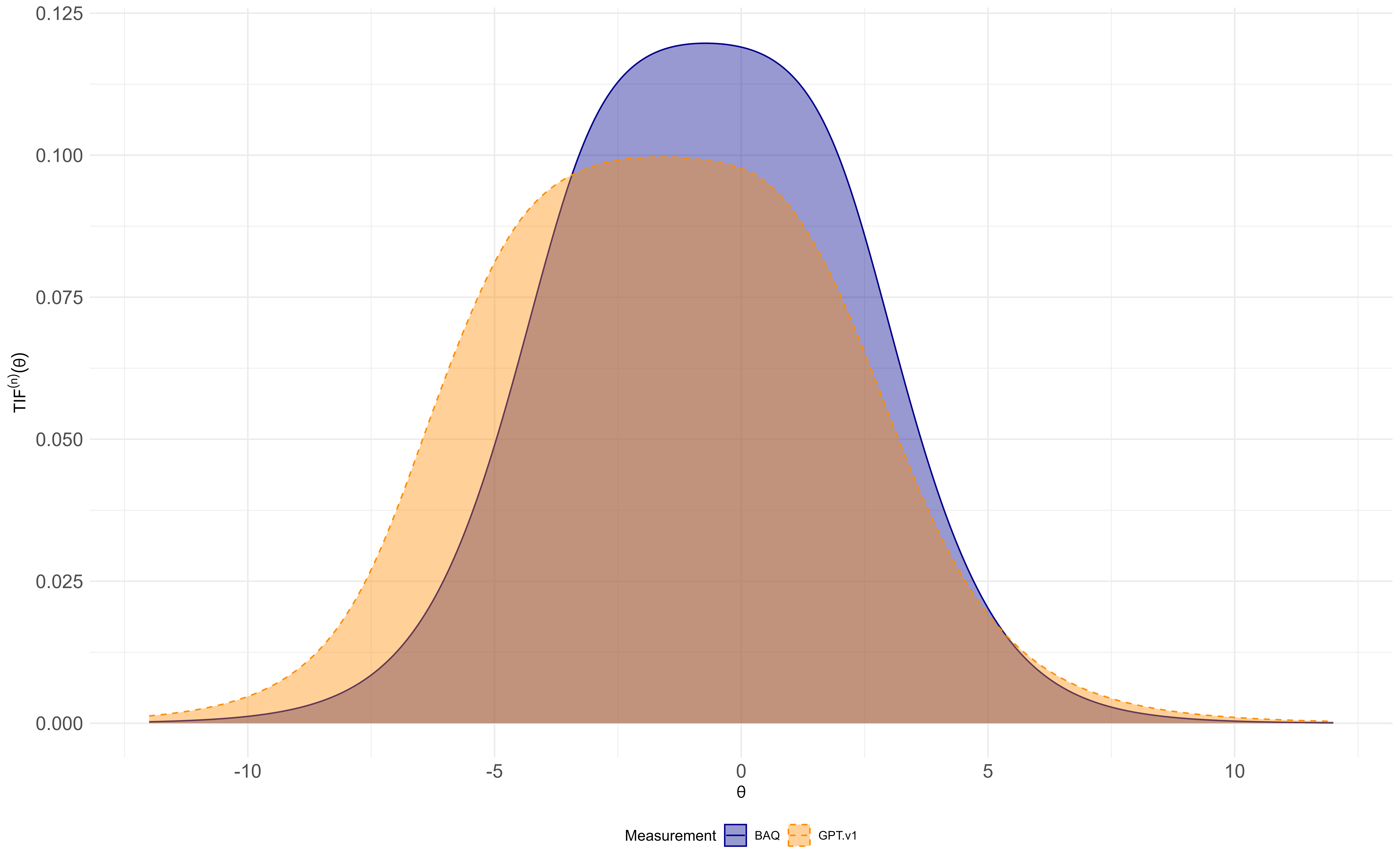}} 
   \caption{Test Information Functions for the original BAQ and the AI-generated GPT.v1 tests.} 
   \label{fig: TIF plots}
\end{figure}

The analogous graphical representations are provided for the TIF (\ref{eq: test information function}) and its normalised version (\ref{eq: weighted test information function}) in Figure \ref{fig: TIF plots}. 
In conclusion, Table \ref{tab: IIFs and TIF overlaps} reports the aforementioned indices, namely: 
\begin{itemize} 
\item the values of the normalising constants (\ref{eq: normalising constant IIF}) evaluated for all three tests (columns $C_{\mathrm{BAQ}}$, $C_{\mathrm{GPT.v1}}$, and $C_{\mathrm{GPT.v2}}$ in Table \ref{tab: IIFs and TIF overlaps}); 
\item the overlap indices for information functions associated with items $\cap(\mathrm{IIF}_{j}^{\mathrm{BAQ}},\mathrm{IIF}_{j}^{\mathrm{GPT.v1}})$ and test $\cap(\mathrm{TIF}^{\mathrm{BAQ}},\mathrm{TIF}^{\mathrm{GPT.v1}})$ (column $\cap(\mathrm{IF})$); 
\item the overlap indices for their scaled versions $\cap(\mathrm{IIF}_{j}^{(n),\mathrm{BAQ}},\mathrm{IIF}_{j}^{(n),\mathrm{GPT.v1}})$ and $\cap(\mathrm{TIF}^{(n),\mathrm{BAQ}},\mathrm{TIF}^{(n),\mathrm{GPT.v1}})$ (column $\cap(\mathrm{IF^{(n)}})$); 
\item the dominance indices (\ref{eq: dominance index}) of the BAQ's normalised IIF over the corresponding function for GPT.v1 (column $\mathrm{DM(BAQ)}$), and vice versa (column $\mathrm{DM(GPT.v1)}$). 
\end{itemize}
 
\begin{table}[!ht]
\centering
\caption{Total IIFs (\ref{eq: normalising constant IIF}), normalised overlaps (\ref{eq: scaled overlap index}) for the original IIFs and their scaled versions, and dominance indices (\ref{eq: dominance index}) between BAQ and GPT.v1. Overlaps are also reported for the standard and normalised test information functions.}
\label{tab: IIFs and TIF overlaps}
\begin{tabular}{clllllll} 
\toprule \\ 
 &
  $C_{\mathrm{BAQ}}$ &
  $C_{\mathrm{GPT.v1}}$ &
  $C_{\mathrm{GPT.v2}}$ &
  $\cap(\mathrm{IF})$ &
  $\cap(\mathrm{IF^{(n)}})$ &
  $\mathrm{Dm(BAQ)}$ &
  $\mathrm{Dm(GPT.v1)}$ \\ 
\doubleRule \\ 
$q_{1}$   & 1.572  & 1.212  & 1.638  & 0.794 & 0.855 & 0.772 & 0.373 \\
$q_{2}$   & 1.596  & 1.087  & 1.835  & 0.751 & 0.841 & 0.771 & 0.388 \\
$q_{3}$   & 2.255  & 2.760  & 3.384  & 0.904 & 0.901 & 0.816 & 0.282 \\
$q_{4}$   & 2.555  & 1.941  & 2.700  & 0.780 & 0.846 & 0.791 & 0.362 \\
$q_{5}$   & 4.381  & 3.281  & 3.089  & 0.701 & 0.779 & 0.878 & 0.343 \\
$q_{6}$   & 2.305  & 1.888  & 3.697  & 0.798 & 0.848 & 0.806 & 0.346 \\
$q_{7}$   & 1.577  & 3.113  & 2.596  & 0.690 & 0.891 & 0.637 & 0.472 \\
$q_{8}$   & 3.285  & 3.603  & 2.810  & 0.770 & 0.751 & 0.841 & 0.408 \\
$q_{9}$   & 3.752  & 3.576  & 3.576  & 0.848 & 0.861 & 0.831 & 0.308 \\
$q_{10}$  & 1.735  & 1.013  & 3.440  & 0.681 & 0.806 & 0.796 & 0.397 \\
$q_{11}$  & 2.427  & 3.059  & 3.917  & 0.870 & 0.861 & 0.814 & 0.325 \\
$q_{12}$  & 2.111  & 2.437  & 3.184  & 0.828 & 0.804 & 0.803 & 0.393 \\
$q_{13}$  & 2.728  & 5.381  & 1.635  & 0.604 & 0.728 & 0.759 & 0.513 \\
$q_{14}$  & 1.935  & 1.208  & 1.794  & 0.702 & 0.814 & 0.800 & 0.385 \\
$q_{15}$  & 1.487  & 2.998  & 2.637  & 0.694 & 0.935 & 0.474 & 0.591 \\
$q_{16}$  & 2.190  & 2.117  & 1.918  & 0.883 & 0.890 & 0.749 & 0.360 \\
$q_{17}$  & 1.925  & 3.293  & 3.014  & 0.734 & 0.857 & 0.732 & 0.411 \\
$q_{18}$  & 2.827  & 2.282  & 2.120  & 0.759 & 0.815 & 0.844 & 0.341 \\ 
\midrule
\textbf{Test} & 42.642 & 46.250 & 48.984 & 0.865 & 0.863 & 
& 
\\ 
\bottomrule 
\end{tabular}%
\end{table}

\section{Discussion}
\label{sec: discussion}

The previous analysis highlights the non-trivial relation between human- and AI-generated questionnaires; despite the high performance shown by the three tests in terms of reliability indices (Section \ref{subsec: reliability}), differences between them already emerged in the exploratory factor analysis (Section \ref{subsec: exploratory factor analysis}). This preliminary evidence mainly involved the less specific AI-generated test GPT.v2; in fact, the strong correspondence between the original BAQ and the more specific version of the AI-generated test (GPT.v1) may suggest that the two questionnaires have comparable structures. While this may hold in terms of reliability, our investigation collected evidence on potential discrepancies in the dimensional structure, the location of difficulty and discrimination parameters, and the quality of information carried by items at different values of the latent trait. 

Starting with the dimensional structure, we saw that the original BAQ produces consistent results in terms of exploratory factor analysis and confirmatory analysis of essential dimensionality. On the other hand, such methods led to identifying a non-trivial dimensional structure for AI-generated questionnaires. The adoption of GRM harmonised such results and produced score estimates for latent traits that are compatible with essential unidimensionality. 

The apparent correspondence between the BAQ and GPT.v1 is not reflected in the estimation of item parameters. In particular, we found that there is no co-monotonicity between the two tests; from Figure \ref{fig: BAQ-GPT.v1 boxplot comparison beta}, we see that the most difficult items in the original BAQ do not correspond to the same difficulty order in the first questionnaire generated by ChatGPT. Analogously, the lack of order-preserving properties arises for discrimination parameters, as it is manifest in Figure \ref{fig: BAQ-GPT.v1 boxplot comparison gamma}. 

A relevant aspect that emerges from our analysis is a decreasing trend for the difficulty in accessing higher response values, which is formalised by moving from higher $\delta$ values for the original BAQ to lower values for the GPT.v1 test. For this type of parameter, we can also consider the less specific GPT.v2 test, as they do not refer to items but only to the ordinal structure of the scale. From Figures \ref{fig: delta posterior distributions} and \ref{fig: boxplot delta posterior distributions}, we see that GPT.v2 shows higher threshold levels than the original BAQ. This phenomenon suggests that distinct factors may affect the ordinal structure represented by difficulty levels, including potential uncertainty or ambiguity in the item formulations from a less specific prompt. To disentangle the aspects related to the body awareness levels under examination and the aforementioned type of ambiguity, we relied on the study of item and test information functions. 

Regarding the accuracy of the measurement instruments at different values of the latent trait, graded response modelling also enables a finer analysis of item and test information functions, which provide a rich description of the items' power in differentiating individuals for all values of the respondent's latent trait estimate. In other words, the $j$-th item information function $IIF_{j}(\vartheta)$ quantifies the accuracy of the information carried by the $j$-th item for each value of the latent trait $\vartheta$. We exploited this description by simultaneously considering a similarity and a divergence measure, namely, the overlap (\ref{eq: scaled overlap index}) and dominance (\ref{eq: dominance index}) indices. From Table \ref{tab: IIFs and TIF overlaps}, we see that similarity and divergence properties may coexist: all the overlap indices 
represent at least $72.8\%$ of the total information for each normalised IIF, which hints at a similarity between corresponding items in the two questionnaires. On the other hand, the dominance indices account for at least $63.7\%$ of the total information in the normalised IIFs for most of the items in the original BAQ test, with the only exception of item $q_{15}$, which is the one with the lowest discriminant power in the original BAQ (\ref{eq: ordered gamma labels by median}). Dually, at most $51.3\%$ of the total information that GPT.v1 encodes in $\mathrm{IIF^{(n)}}_{j}$ dominates over the corresponding term in the BAQ, except when $j = 15$. Dealing with normalised functions, these results show that the information in the original BAQ is more concentrated and 
more accurately characterises the latent traits in a given range, while the AI-generated questionnaire GPT.v1 is less specific as it spreads the item information over a larger range. This aspect represents an instance of epistemic uncertainty in results generated by artificial agents \cite{angelelli2024representations}, which, in our context, may affect the construct validity of the latent trait measurement. 

The connection between information and discrimination power hinted at by the anomalous behaviour of the item $q_{15}$ is also observed by considering an association index between the estimated parameters $\hat{\gamma}$ and the overlap or dominance indices. In this sense, we can distinguish the potential effects of the different item characteristics on information functions. First, the discrimination parameter has a major influence on the total IIF: specifically, the item ordering induced by $\hat{\gamma}_{j}$ in (\ref{eq: ordered gamma labels by median}) is in line with the one induced by the total IIFs shown in Table \ref{tab: IIFs and TIF overlaps}, with a single minor difference represented by the transposition $(1,14)$ in $\sigma_{\mathrm{GPT.v1}}$. Furthermore, we observe a high positive association between the ranks of the vector of $\hat{\gamma}^{\mathrm{BAQ}}$ and the dominance indices $\mathrm{Dm}(\mathrm{IIF^{(n),BAQ}},\mathrm{IIF^{(n),GPT.v1}})$ in terms of the Spearman rank correlation $\varrho_{S}=0.763$, as well as a strong negative association between the ranks of $\hat{\gamma}^{\mathrm{BAQ}}$ and the dominance index $\mathrm{Dm}(\mathrm{IIF^{(n),GPT.v1}},\mathrm{IIF^{(n),BAQ}})$, with Spearman rank correlation $\varrho_{S}=-0.517$. This observation is in line with Figure \ref{fig: IIF normalised plots}, where we can see that the original BAQ is generally more informative when $\vartheta$ lies in the central range, while the AI-generated test GPT.v1 is slightly more informative for extreme values of the latent trait. 

The difficulty parameters $\beta$ do not have such effects, but they may have a partial influence on item informativeness about higher values of the latent trait. In more detail, from Figure \ref{fig: IIF normalised plots}, we can see that the $j$-th item in the BAQ tends to be more informative about higher values of $\vartheta$ when $\hat{\beta}_{j}$ is large or, referring to (\ref{eq: ordered beta labels by median}), has a higher position in $\pi_{\mathrm{BAQ}}$ compared to the one in $\pi_{\mathrm{GPT.v1}}$. At the test level, this behaviour is manifest for both the original test information function (Figure \ref{fig: TIF plot}) and its normalised versions (Figure \ref{fig: wTIF plot}). 

These observations suggest that the AI-generated measurement instruments may underline different dimensional structures that emerge under appropriate statistical methods, while the original BAQ questionnaire is more consistent in this sense. The test generated by ChatGPT also affects the item characteristics, even when an apparent correspondence with the items in the original test is observed. This aspect is more manifest in the GPT.v2 questionnaire: even if there is no strict correspondence with the items in the original BAQ, we can compare the contributions to test difficulty from the set of items (the ordered list of $\beta$ parameters) and the ordinal structure ($\delta$ parameters), as they do not depend on such correspondence. From Figure \ref{fig: sorted boxplot beta posterior distributions}, we see that GPT.v2 tends to simplify items; on the other hand, its low specificity may generate ambiguity and increase the difficulty in accessing higher response levels exploring the awareness trait, as it is shown in Figure \ref{fig: boxplot delta posterior distributions}. In this way, we can appreciate the effects of distinct measurement instruments on the decomposition of the overall item difficulty terms $\beta_{j,h}$. Such effects, along with the implications of the lack of co-monotonicity discussed above, are reflected in the way the items and the overall test convey information about the range of latent traits.

\section{Conclusion} 
\label{sec: conclusion}

In this work, we started an exploration of tests generated by large language models, specifically ChatGPT, in comparison with validated tests produced by human experts. This investigation highlighted that, despite their good performance in terms of reliability, some relevant differences emerge when looking at the test dimensionality and information provided by the tests about the respondents' latent traits. The identification of these differences is relevant in deriving and interpreting the results of measurement processes involving individuals' perception; as a 
case study, we explored the body awareness construct starting with a validated questionnaire, and special attention was given to the wording of the prompts. The study relied on the tools provided by IRT modelling, which overcomes some limitations of score-based approaches and allows for a deeper analysis of the underlying factors that drive the individuals' responses. Furthermore, we were able to identify differences in the ranking of items within the questionnaire, linking them to the information carried across different values of the latent traits. 

The proposed methodology can be further developed in several directions. First, the functional approach relying on (item or test) information function, as well as posterior distributions from Bayesian estimation, suggests extending overlap and dominance indices studied in this work to \emph{information divergences}, such as Kullback-Leibler divergence, or distance between distributions, such as Wasserstein. The former is a natural measure of discrepancy between distributions obtained from different tests or item versions based on entropy, which can be used to explore geometric criteria to study transformations of functions characterising distributions \citep{angelelli2021entropy}; the latter has already proven useful for model comparison and selection in the context of SAFE (Sustainable, Accurate, Fair, and Explainable) AI \citep{spelta2024evaluating}. In parallel, co-monotonicity (Section \ref{subsec: lack of co-monotonicity}) relates to deviations from rank similarity between the original and the AI-generated questionnaires. In this perspective, further analysis could adapt Rank Graduation Accuracy \citep{giudici2025rga}---a measure suited to AI explainability assessment \citep{babaei2025rank}---and related indices \citep{angelelli2024robust} to explore whether AI-generated test adaptations with an appropriate prompt may be informative about original items, even considering SAFE dimensions \citep{giudici2024explainable}. However, this analysis would require a repeated-measures design in the data collection. More generally, given the exploratory nature of the present study and the relatively small sample size for each questionnaire version, further research based on an extended sampling process is required to assess the generalisability of the preliminary findings reported herein. 

Considering covariates and extending the IRT model to account for them may better characterise latent trait variability. These might be reflected in the dimensional structure and in the different ways AI-generated questionnaires are perceived; the assessment of such differences could be assessed, for example, by differential item functioning since this notion is well suited to identifying item bias \citep{penfield20065}. We can further extend IRT models by considering mIRT \citep{Reckase2009}, or structural equation modelling, which can account for higher model complexity in terms of the number of latent constructs and relationships between them. 

Finally, this work focused on a case study related to the psychological domain, with particular relevance to the study of subjects' mental health and wellness. Other factors related to social dimensions could be explored, with particular focus on the detection and perception of stereotypes, as biases in data used to train large language models may encompass such stereotypes \citep{Abid_2021} and affect the fairness of the LLM outputs \citep{rozado2023political}. This approach could support the aware adoption of AI-generated measurements, acknowledging the latent differences with their human-generated counterparts and the role of appropriate prompt design by experts in social research.

\bibliographystyle{plainnat}
\bibliography{bibliography.bib}  

\newpage
\begin{appendices}

\section{Summary of GRM estimates from MCMC sampling} 
\label{app: summary of GRM estimates from MCMC sampling} 

\begin{table}[!ht]
\centering
\caption{Summary of MCMC sampling for difficulty parameters.}
\label{tab: beta MCMC summary}
\begin{tabular}{clllllllll} 
\toprule \\ 
 &
  \multicolumn{3}{c}{\textbf{BAQ}} &
  \multicolumn{3}{c}{\textbf{GPT.v1}} &
  \multicolumn{3}{c}{\textbf{GPT.v2}} \\ 
\doubleRule \\ 
\multicolumn{1}{c}{} &
  \multicolumn{1}{c}{Mean} &
  \multicolumn{1}{c}{SD} &
  \multicolumn{1}{c}{Median} &
  \multicolumn{1}{c}{Mean} &
  \multicolumn{1}{c}{SD} &
  \multicolumn{1}{c}{Median} &
  \multicolumn{1}{c}{Mean} &
  \multicolumn{1}{c}{SD} &
  \multicolumn{1}{c}{Median} \\ 
\doubleRule \\ 
$\beta_{1}$  & -2.110 & 1.103 & -2.250 & -1.254 & 1.086 & -1.231 & -1.271 & 1.339 & -1.457 \\
$\beta_{2}$  & -1.242 & 1.080 & -1.389 & -0.624 & 1.090 & -0.599 & -0.732 & 1.334 & -0.923 \\
$\beta_{3}$  & -1.544 & 1.067 & -1.693 & -2.082 & 1.052 & -2.056 & -1.484 & 1.323 & -1.670 \\
$\beta_{4}$  & -0.366 & 1.051 & -0.521 & -0.836 & 1.052 & -0.812 & -1.894 & 1.334 & -2.077 \\
$\beta_{5}$  & -0.733 & 1.043 & -0.893 & -2.010 & 1.046 & -1.981 & -2.762 & 1.341 & -2.951 \\
$\beta_{6}$  & -0.987 & 1.061 & -1.140 & -1.755 & 1.064 & -1.730 & -2.914 & 1.345 & -3.104 \\
$\beta_{7}$  & -1.126 & 1.083 & -1.274 & -2.027 & 1.048 & -2.005 & -2.900 & 1.350 & -3.082 \\
$\beta_{8}$  & -1.047 & 1.052 & -1.202 & -2.826 & 1.061 & -2.799 & -2.043 & 1.334 & -2.229 \\
$\beta_{9}$  & -0.898 & 1.048 & -1.055 & -1.488 & 1.034 & -1.465 & -1.973 & 1.328 & -2.165 \\
$\beta_{10}$ & -0.331 & 1.068 & -0.485 & -1.121 & 1.104 & -1.093 & -1.913 & 1.329 & -2.105 \\
$\beta_{11}$ & -0.042 & 1.054 & -0.204 & -0.952 & 1.032 & -0.924 & -2.848 & 1.343 & -3.038 \\
$\beta_{12}$ & 0.373  & 1.057 & 0.215  & -1.161 & 1.046 & -1.144 & -3.155 & 1.350 & -3.341 \\
$\beta_{13}$ & -1.083 & 1.057 & -1.238 & -3.190 & 1.062 & -3.161 & -0.772 & 1.339 & -0.959 \\
$\beta_{14}$ & -1.352 & 1.072 & -1.504 & -2.204 & 1.100 & -2.176 & -1.446 & 1.336 & -1.630 \\
$\beta_{15}$ & -0.738 & 1.078 & -0.889 & -1.184 & 1.035 & -1.157 & -2.527 & 1.346 & -2.714 \\
$\beta_{16}$ & -0.327 & 1.055 & -0.482 & -0.648 & 1.048 & -0.626 & -1.363 & 1.336 & -1.546 \\
$\beta_{17}$ & -1.100 & 1.070 & -1.249 & -2.231 & 1.050 & -2.206 & -2.484 & 1.339 & -2.670 \\
$\beta_{18}$ & -1.848 & 1.068 & -1.998 & -2.958 & 1.082 & -2.932 & -2.204 & 1.344 & -2.387 \\ 
\bottomrule 
\end{tabular}%
\end{table}

\begin{table}[!ht]
\centering
\caption{Summary of MCMC sampling for discrimination parameters.}
\label{tab: gamma MCMC summary}
\begin{tabular}{clllllllll} 
\toprule 
 &
  \multicolumn{3}{c}{\textbf{BAQ}} &
  \multicolumn{3}{c}{\textbf{GPT.v1}} &
  \multicolumn{3}{c}{\textbf{GPT.v2}} \\ 
\doubleRule \\ 
\multicolumn{1}{c}{} &
  \multicolumn{1}{c}{Mean} &
  \multicolumn{1}{c}{SD} &
  \multicolumn{1}{c}{Median} &
  \multicolumn{1}{c}{Mean} &
  \multicolumn{1}{c}{SD} &
  \multicolumn{1}{c}{Median} &
  \multicolumn{1}{c}{Mean} &
  \multicolumn{1}{c}{SD} &
  \multicolumn{1}{c}{Median} \\ 
\doubleRule \\ 
$\gamma_{1}$  & 0.728 & 0.132 & 0.719 & 0.571 & 0.100 & 0.564 & 0.731 & 0.112 & 0.724 \\
$\gamma_{2}$  & 0.735 & 0.126 & 0.727 & 0.532 & 0.089 & 0.525 & 0.786 & 0.123 & 0.777 \\
$\gamma_{3}$  & 0.920 & 0.158 & 0.909 & 0.963 & 0.160 & 0.951 & 1.159 & 0.191 & 1.145 \\
$\gamma_{4}$  & 0.993 & 0.163 & 0.982 & 0.773 & 0.131 & 0.763 & 1.005 & 0.160 & 0.994 \\
$\gamma_{5}$  & 1.401 & 0.262 & 1.379 & 1.077 & 0.195 & 1.060 & 1.096 & 0.197 & 1.080 \\
$\gamma_{6}$  & 0.933 & 0.169 & 0.920 & 0.758 & 0.130 & 0.749 & 1.229 & 0.227 & 1.212 \\
$\gamma_{7}$  & 0.730 & 0.129 & 0.720 & 1.039 & 0.178 & 1.025 & 0.982 & 0.182 & 0.968 \\
$\gamma_{8}$  & 1.164 & 0.197 & 1.149 & 1.145 & 0.225 & 1.124 & 1.032 & 0.173 & 1.019 \\
$\gamma_{9}$  & 1.269 & 0.220 & 1.252 & 1.140 & 0.206 & 1.121 & 1.199 & 0.198 & 1.185 \\
$\gamma_{10}$ & 0.774 & 0.127 & 0.765 & 0.509 & 0.088 & 0.502 & 1.171 & 0.191 & 1.157 \\
$\gamma_{11}$ & 0.962 & 0.167 & 0.949 & 1.029 & 0.176 & 1.014 & 1.276 & 0.236 & 1.257 \\
$\gamma_{12}$ & 0.881 & 0.156 & 0.867 & 0.890 & 0.147 & 0.879 & 1.124 & 0.210 & 1.106 \\
$\gamma_{13}$ & 1.036 & 0.182 & 1.022 & 1.478 & 0.300 & 1.451 & 0.730 & 0.118 & 0.722 \\
$\gamma_{14}$ & 0.835 & 0.153 & 0.823 & 0.573 & 0.105 & 0.564 & 0.776 & 0.130 & 0.767 \\
$\gamma_{15}$ & 0.705 & 0.128 & 0.695 & 1.015 & 0.173 & 1.002 & 0.994 & 0.177 & 0.980 \\
$\gamma_{16}$ & 0.900 & 0.152 & 0.889 & 0.815 & 0.140 & 0.804 & 0.811 & 0.131 & 0.801 \\
$\gamma_{17}$ & 0.832 & 0.149 & 0.821 & 1.080 & 0.190 & 1.064 & 1.084 & 0.183 & 1.069 \\
$\gamma_{18}$ & 1.063 & 0.186 & 1.049 & 0.856 & 0.157 & 0.843 & 0.861 & 0.138 & 0.852 \\ 
\bottomrule 
\end{tabular}%
\end{table}

\begin{table}[!ht]
\centering
\caption{Summary of MCMC sampling for response level thresholds.}
\label{tab: delta MCMC summary}
\begin{tabular}{clllllllll} 
\toprule \\ 
 & \multicolumn{3}{c}{\textbf{BAQ}} & \multicolumn{3}{c}{\textbf{GPT.v1}} & \multicolumn{3}{c}{\textbf{GPT.v2}} \\ 
\doubleRule \\ 
 &
  \multicolumn{1}{c}{Mean} &
  \multicolumn{1}{c}{SD} &
  \multicolumn{1}{c}{Median} &
  \multicolumn{1}{c}{Mean} &
  \multicolumn{1}{c}{SD} &
  \multicolumn{1}{c}{Median} &
  \multicolumn{1}{c}{Mean} &
  \multicolumn{1}{c}{SD} &
  \multicolumn{1}{c}{Median} \\ 
\doubleRule \\ 
$\delta_{1}$ & -2.254    & 1.042    & -2.085    & -3.004     & 1.067     & -3.014     & -2.329     & 1.323     & -2.155     \\
$\delta_{2}$ & -1.191    & 1.022    & -1.013    & -1.432     & 1.012     & -1.458     & -1.310     & 1.304     & -1.119     \\
$\delta_{3}$ & -0.309    & 1.018    & -0.129    & -0.425     & 0.997     & -0.461     & -0.303     & 1.297     & -0.107     \\
$\delta_{4}$ & 0.628     & 1.021    & 0.803     & 0.640      & 0.996     & 0.608      & 0.810      & 1.298     & 1.000      \\
$\delta_{5}$ & 1.560     & 1.033    & 1.730     & 1.913      & 1.010     & 1.882      & 1.841      & 1.307     & 2.029      \\
$\delta_{6}$ & 2.737     & 1.060    & 2.896     & 3.347      & 1.045     & 3.316      & 3.016      & 1.325     & 3.213     \\ 
\bottomrule 
\end{tabular}%
\end{table} 

\end{appendices}

\end{document}